\RequirePackage{fix-cm} % Fix LaTeX2e bugs.
\documentclass[a4paper, twoside, reqno, dvips, 12pt]{amsart}
\usepackage{fixltx2e}   % Fix LaTeX2e bugs.

\usepackage{etex}

\usepackage[latin1]{inputenc}
\usepackage[T1]{fontenc}

\usepackage{esint}
\usepackage{dsfont}
\usepackage{xspace}
\usepackage{amsgen}
\usepackage{amsthm}
\usepackage{amssymb}
\usepackage{amsmath}
\usepackage{wasysym}
\usepackage{upgreek}
\usepackage{amsfonts}
\usepackage{stmaryrd}
\usepackage{mathtools}

\usepackage{relsize}
\usepackage[mathcal, mathscr]{euscript}

\usepackage{mathrsfs}
\DeclareMathAlphabet{\mathscrbf}{OMS}{mdugm}{b}{n}

\usepackage{a4wide}

\usepackage[dvipsnames, table]{xcolor}
\definecolor{bckg}{RGB}{20.8, 20.8, 20.8}
\definecolor{oneblue}{rgb}{0.0, 0.0, 0.85}
\definecolor{Lightblue}{RGB}{214, 214, 214}
\definecolor{bluepigment}{rgb}{0.2, 0.2, 0.6}
\definecolor{charcoal}{rgb}{0.21, 0.27, 0.31}
\definecolor{denimblue}{rgb}{0.08, 0.38, 0.74}
\definecolor{Lightgray}{rgb}{0.89, 0.89, 0.89}
\definecolor{darkgrey}{rgb}{0.273, 0.281, 0.30}
\definecolor{darkelectricblue}{rgb}{0.33, 0.41, 0.47}

\usepackage[sort&compress, comma, square, numbers]{natbib}

\usepackage{psfrag}
\usepackage{graphicx}
\usepackage{subfigure}
\usepackage{morefloats}
\usepackage{indentfirst}

\usepackage{acronym}
\usepackage{microtype}
\usepackage[labelsep=period,%
            labelfont={bf,sf,color=bluepigment},%
            justification=raggedright]{caption}

\usepackage[perpage, symbol]{footmisc}

\usepackage[usenames, dvipsnames, pdf]{pstricks}
\usepackage{epsfig}
\usepackage{pst-grad} % For gradients
\usepackage{pst-plot} % For axes

\usepackage[colorlinks,
           urlcolor=oneblue,
           linkcolor=denimblue,
           citecolor=NavyBlue,
           bookmarksopen=false,
           pdfpagemode=UseNone,
           pagebackref]{hyperref}

\usepackage[explicit]{titlesec}

\titleformat{\section}[block]
  {\color{NavyBlue}\Large\sffamily\bfseries}
  {}
  {0.0em}
  {\colorbox{bckg!5}{\strut\parbox{\dimexpr\linewidth-2\fboxsep\relax}{\thesection. #1}}}
  [\vspace*{0.33em}]

\titleformat{name=\section,numberless}[block]
  {\color{NavyBlue}\Large\sffamily\bfseries}
  {}
  {0.0em}
  {\colorbox{bckg!5}{\strut\parbox{\dimexpr\linewidth-2\fboxsep\relax}{#1}}}
  [\vspace*{0.33em}]

\titleformat{\subsection}
  {\color{NavyBlue}\large\sffamily\bfseries}
  {}
  {0.0em}
  {\colorbox{bckg!5}{\parbox{\dimexpr\linewidth-2\fboxsep\relax}{\thesubsection. #1}}}
  [\vspace*{0.33em}]

\titleformat{name=\subsection,numberless}
  {\color{NavyBlue}\Large\sffamily\bfseries}
  {}
  {0em}
  {\colorbox{bckg!5}{\parbox{\dimexpr\linewidth-2\fboxsep\relax}{#1}}}
  [\vspace*{0.33em}]

\titleformat{\subsubsection}
  {\color{bluepigment}\sffamily\normalsize\bfseries}
  {\thesubsubsection}
  {0.5em}
  {#1}
  [\vspace*{0.33em}]

\titleformat{\paragraph}[runin]
  {\color{bluepigment}\sffamily\small\bfseries}
  {}
  {0em}
  {#1}

\titlespacing{\section}{0.0em}{1.5em plus 2pt minus 2pt}%
{1.0em plus 2pt minus 2pt}[0em]
\titlespacing{\subsection}{0.5em}{1.5em plus 2pt minus 2pt}%
{1.0em}[0em]
\titlespacing{\subsubsection}{0.5em}{1.5em plus 2pt minus 2pt}%
{1.0em plus 2pt minus 2pt}[0em]

\usepackage{titletoc}

\contentsmargin{0.5em}
\newlength{\tocsep} 
\titlecontents{section}[\tocsep]
  {\addvspace{10pt}\bfseries\sffamily}
  {\contentslabel[\thecontentslabel]{\tocsep}}
  {}
  {\ \titlerule*[0.75pc]{.}\ \thecontentspage}
  []
\titlecontents{subsection}[\tocsep]
  {\addvspace{8pt}\sffamily}
  {\contentslabel[\thecontentslabel]{\tocsep}}
  {}
  {\ \titlerule*[0.5pc]{.}\ \thecontentspage}
  []
\titlecontents*{subsubsection}[\tocsep]
  {\addvspace{2pt}\footnotesize\sffamily}
  {}
  {}
  {\ \titlerule*[0.35pc]{.}\ \thecontentspage}
  [\\*]

\makeatletter
\def\@setauthors{%
  \begingroup
  \def\thanks{\protect\thanks@warning}%
  \trivlist
  \centering\footnotesize \@topsep30\p@\relax
  \advance\@topsep by -\baselineskip
  \item\relax
  \author@andify\authors
  \def\\{\protect\linebreak}%
  \textsc{\normalsize\textcolor{darkelectricblue}{\authors}}%
  \ifx\@empty\contribs
  \else
    ,\penalty-3 \space \@setcontribs
    \@closetoccontribs
  \fi
  \endtrivlist
  \endgroup
}
\def\@settitle{\begin{center}%
  \baselineskip14\p@\relax
    \bfseries
    \textsc{\Large\textcolor{charcoal}{\@title}}
  \end{center}%
}
\makeatother

\usepackage{enumitem}
\setlist[description]{%
  topsep=30pt,               % space before start / after end of list
  itemsep=5pt,               % space between items
  font={\bfseries\sffamily\color{NavyBlue}}, % if colour is needed
}

\usepackage{fancyhdr}
\usepackage{lastpage}

\newcommand*\Title{\textcolor{bluepigment}{Cascade in the modified KdV equation}}
\newcommand*\Authors{\textcolor{bluepigment}{D.~Dutykh \& E.~Tobisch}}
\newcommand*{\plogo}{\textcolor{gray}{{\texttt{arXiv.org} / \textsc{hal}}}} % Generic publisher logo

\numberwithin{equation}{section}

\newcommand{\up}[1]{$\,^{\mathrm{\small\textsf{#1}}}$} % \up command from French babel

\newcommand{\R}{\mathbb{R}}
\newcommand{\Z}{\mathbb{Z}}
\newcommand{\E}{\mathcal{E}}

\newcommand{\ud}{\mathrm{d}}
\newcommand{\ui}{\mathrm{i}}

\newcommand{\eps}{\varepsilon}
\renewcommand{\O}{\mathcal{O}}

\newcommand{\pd}[2]{\frac{\partial\, #1}{\partial\/ #2}}

\acrodef{kdv}[KdV]{Korteweg--de Vries}
\acrodef{wtt}[WTT]{Wave Turbulence Theory}
\acrodef{mi}[MI]{Modulational Instability}
\acrodef{nls}[NLS]{Nonlinear Schr\"{o}dinger}
\acrodef{ode}[ODE]{Ordinary Differential Equation}
\acrodef{mnls}[mNLS]{modified Nonlinear Schr\"{o}dinger}
\acrodef{mkdv}[mKdV]{modified Korteweg--de Vries}
\acrodef{gkdv}[gKdV]{generalized Korteweg--de Vries}
\acrodef{icem}[ICEM]{Increment Chain Equation Method}

\newcommand{\ie}{\emph{i.e.}\xspace}
\newcommand{\eg}{\emph{e.g.}\xspace}

\begin{document}

\title[\Title]{Direct dynamical energy cascade in the modified KdV equation}

\author[D.~Dutykh]{Denys Dutykh}
\address{LAMA, UMR 5127 CNRS, Universit\'e de Savoie, Campus Scientifique, 73376 Le Bourget-du-Lac Cedex, France}
\email{Denys.Dutykh@univ-savoie.fr}
\urladdr{http://www.denys-dutykh.com/}

\author[E.~Tobisch]{Elena Tobisch$^*$}
\address{Institute for Analysis, Johannes Kepler University, Linz, Austria}
\email{Elena.Tobisch@jku.at}
\urladdr{http://www.dynamics-approx.jku.at/lena/}
\thanks{$^*$ Corresponding author}

%%% ------------------------------------------------------------------------ %%%

\begin{titlepage}
\setcounter{page}{1}
\thispagestyle{empty} % Remove page numbering on this page
\noindent
{\Large Denys \textsc{Dutykh}}\\
{\it\textcolor{gray}{CNRS--LAMA, University of Savoie, France}}\\[0.02\textheight]
{\Large Elena \textsc{Tobisch}}\\
{\it\textcolor{gray}{Johannes Kepler University, Linz, Austria}}\\[0.16\textheight]

\vspace*{1cm}

\colorbox{Lightblue}{
  \parbox[t]{1.0\textwidth}{
    \centering\huge\sc
    \vspace*{0.7cm}

    \textcolor{bluepigment}{Direct dynamical energy cascade in the modified KdV equation}

    \vspace*{0.7cm}
  }
}

\vfill % Whitespace between the title block and the publisher

\raggedleft     % Right-align all text
{\large \plogo} % Publisher and logo
\end{titlepage}

%%% ------------------------------------------------------------------------ %%%

\newpage
\maketitle
\thispagestyle{empty}

\begin{abstract}

In this study we examine the energy transfer mechanism during the nonlinear stage of the \acf{mi} in the \acf{mkdv} equation. The particularity of this study consists in considering the problem essentially in the Fourier space. A dynamical energy cascade model of this process originally proposed for the focusing \acs{nls}-type equations is transposed to the \acs{mkdv} setting using the existing connections between the \acs{kdv}-type and \acs{nls}-type equations. The main predictions of the $D$-cascade model are outlined and  validated by direct numerical simulations of the \acs{mkdv} equation using the pseudo-spectral methods. The nonlinear stages of the \acs{mi} evolution are also investigated for the \acs{mkdv} equation.

\bigskip\bigskip
\noindent \textbf{\keywordsname:} Modulational Instability; energy cascade; \acl{kdv} equation; modified KdV equation; NLS equation. \\


\end{abstract}

\newpage
\tableofcontents
\thispagestyle{empty}
\newpage

%%% ------------------------------------------------------------------------ %%%

\section{Introduction}

Nonlinear  wave systems occur in numerous physical areas from optics to fluid mechanics, from astronomy to geophysics, and  one of the most important issues regarding these systems is a description of its energy behavior. One of the most beautiful examples to illustrate this point, is the hypothesis of  Kolmogorov on the form of energy spectrum in systems with \emph{strong} turbulence in which the energy spectrum is supposed to have the universal form $\E(\ell) \sim \ell^{-5/3}$ where $\ell$ is the size of the eddy, \cite{Kolmogorov1991}. In the kinetic \emph{weak}  (or wave) turbulence theory (WTT) dispersive waves play now the role of eddies, and the energy spectrum is again power law $k^{-\alpha}$ where $k$ is the wave length (if dispersion function $\omega \sim k^{\beta}$) and $\alpha$ is not universal any more, \cite{Zakharov1992}.

The kinetic \acs{wtt} is an asymptotic theory which is working for very small nonlinearity $0 < \eps \lessapprox 0.01$, where small parameter $\eps$ is usually taken as a product of wave amplitude with wave number, $\eps = Ak$. The smallness of $\eps$ is very important while the kinetic \acs{wtt} is essentially based on the following \emph{assumption}: time scales for $3$--, $4$--, $\ldots$, $s$--wave resonances are separated and can be studied independently. This assumption breaks at about $\eps \approx 0.1$, \eg \cite{Annenkov2006a}. On the other hand, usual laboratory experiments and numerical simulations are performed for  $\eps \approx 0.1 \div 0.4$ while  for a smaller  $\eps \sim 0.01$ corresponding time scales  are too long and kinetic energy cascades  can not be observed in an experiment at the present stage of technical facilities, \cite{Kartashova2013, Newell2011}.

A new model (hereafter referred to as $D$-model) for the formation of the energy spectrum has been developed by E.~\textsc{Kartashova} (2012) in \cite{Kartashova2012}; the model can be applied for describing nonlinear wave systems with nonlinearity parameter of the order of $\eps\ \sim\ 0.1 \div 0.4$ and wave systems with narrow frequency band excitation. Basic physical mechanism responsible for the formation of the energy spectrum in this model is not a common $s$--wave resonance but the modulation instability, and the main assumption of the model is that energy cascade is formed by the most unstable modes in the system, \ie modes with maximum increment of instability. In \cite{Kartashova2012}, the \acf{icem}  was developed for computing dynamical energy spectrum in the systems possessing modulation instability, and applied for the focusing \acs{nls} and \acs{mnls}, with different levels on nonlinearity, \cite{Kartashova2012a, Kartashova2014}.

The \acs{nls} is a very attractive equation because of its integrability, but unfortunately it gives sometimes not good enough description of the observed physical effects. For instance,  modulation instability was discovered in laboratory experiments with water waves and explained by \textsc{Benjamin} \& \textsc{Feir} (1967) \cite{Benjamin1967a}, as instability of a narrow wave packet in the framework of the \acs{nls}.  However, numerical simulations with \acs{nls} demonstrate a symmetric energy cascade in the Fourier space while energy cascade experimentally observed in a water tank, is asymmetric. To cope with this problem, it is necessary to introduce various modifications to \acs{nls}, \eg \cite{Dysthe1979, Hogan1985}. These modifications allow for the realistic values of small parameter, $0 < \eps_{\mathrm{real}} \sim 0.1 \div 0.4$, and are more suitable for modeling real physical phenomena.

The \acf{kdv} equation along with its various modifications is another widely used model equation describing long waves, \ie the region of small wave vectors $kd\ll 1$, $d$ being the mean water depth. This equation does not have the restrictive assumption of narrow spectrum as the \acs{nls} equation. However, many modifications of \acs{kdv} are integrable which is a strong mathematical property. So it is not surprising that the \acs{kdv} equation has found many applications in different fields of Physics such as the shallow water wave dynamics \cite{Zabusky1971, Dullin2004}, internal waves in two-component fluids \cite{Grimshaw2001} and acoustic waves in plasmas \cite{Schamel1973}. Our motivation to study the \acs{kdv}-family of equations comes mainly from the numerous real-world applications that it can cover.

Though KdV does not have modulation instability \cite{Ablowitz1981}, its various modifications do possess this property, under certain conditions. Thus, perturbations of a quasi-periodical wave train with small amplitudes in the generalized KdV equations with  nonlinearity of the form $(u^{p+1})_x$
\begin{equation}\label{gKdV}
  \mathrm{gKdV}(u_{\pm}) \doteq u_t + u_{xxx} + (u^{p+1})_x = 0\ ,
\end{equation}
are modulationally stable if $p < 2$, while they are modulationally unstable if $p > 2$, \cite{Haragus2008}. A more general version of this result allowing for nonlocal dispersion can be found in \cite{Johnson2013}. However, all these results do not allow to obtain a nice analytical representation for the instability interval as in \cite{Benjamin1967a}, and a numerical study is unavoidable.

Another reference point important for our study of mKdV is the following remarkable feature of this equation: it can be reduced, under certain conditions, to the mNLS where the MI can be studied by analytically. This reduction can be made by the variational methods \cite{Whitham1967a} or by standard asymptotical approach as in \cite{Grimshaw2001}.

Our aim in this paper is to study a particular case with $p = 2$ and one space dimension --- the so-called \acf{mkdv} equation
\begin{equation}\label{mKdV}
  \mathrm{mKdV}(u_{\pm}) \doteq u_t + u_{xxx} \pm 6u^2u_x = 0\ ,
\end{equation}
with $u$ being a real-valued scalar function, $x$ and $t$ are space and time variables consequently, and the subscripts denote the  corresponding partial derivatives. As a starting point for our simulations aiming to study the \acs{mi} in the \acs{mkdv}, i.e. $p\ =\ 2$ in \eqref{gKdV}, we use the estimates obtained in \cite{Haragus2008} by combination of analytical results and numerical estimates, namely that for $p=2$, the wave is spectrally stable for all wave vectors $0\ <\ k^2\ <\ 2$.

It is also shown in \cite{Grimshaw2001} that wave packets are unstable only for a positive sign of the coefficient of the cubic nonlinear term in \eqref{mKdV}, and for a high carrier frequency. Being interested in modulation instability, we restrict the study further on the case of focusing mKdV equation:
\begin{equation}\label{mKdV+}
  \mathrm{mKdV}(u_{+}) \doteq u_t + u_{xxx} + 6u^2u_x = 0.
\end{equation}
In the present paper we aim to study in detail formation and properties of the  direct  $D$-cascade in the frame of \acs{kdv} equation \eqref{mKdV+}.

The present manuscript is organized as follows. In Section~\ref{s:cascade} we give a sketch of a $D$-cascade formation for this equation and formulate the properties of the cascade and its spectra which should be verified numerically. In Section~\ref{s:numerics} we describe shortly our numerical approach and present results of our numerical simulations. Finally, the main conclusions of this study are briefly formulated in Section~\ref{sec:disco}.

%%%%%%%%%%%%%%%%%%%%%%%%%%%%%%%%%%%%%%%%%%%%%%%%%%%%%%%%%%%%%%%%%%%%%%

\section{$D$-cascade in the model equation}\label{s:cascade}

The main effect of the \acf{mi} is the disintegration of periodic wavetrains into side bands. \textsc{Benjamin} \& \textsc{Feir} (1967) \cite{Benjamin1967a} showed that there is a connection between the frequencies, wavenumber and amplitudes of unstable modes in the framework of the focusing ($+$) \acf{nls} equation, which reads after a proper re-scaling:
\begin{equation*}
  \mathrm{NLS}(v_{\pm}) \doteq \ui v_t + v_{xx} \pm |v|^2v = 0
\end{equation*}
Namely, they computed the instability interval in the form
\begin{equation}\label{InstInter}
  0\ <\ {\Delta \omega} \bigr/ {A k\omega}\ \le\ \sqrt{2},
\end{equation}
where $\omega(k)$ is the linear dispersion relation, $k$ is the wavenumber and $A$ is the amplitude of the Fourier mode $\omega$. Quantity $\Delta\omega$ is the distance between the parent mode and its side band. It was also shown in \cite{Benjamin1967a} that the most unstable mode satisfies the following relation:
\begin{equation}\label{eq:max}
  {\Delta \omega} \bigr/ {A k\omega}\ =\ 1.
\end{equation}
The use of two assumptions - a) an energy cascade is formed by the most unstable modes, and b) the energy fraction $p$ (called cascade intensity) transported from one cascading mode to the next one is constant, allows to construct and to solve an approximate ordinary differential equation for computing amplitudes of cascading modes, \cite{Kartashova2012}. The first constitutive assumption was  inspired by the well-known hypothesis of O. Phillips while the second - by numerous experimental studies of water waves, \eg \cite{Xia2010}.

The amplitude of the $n$-s mode in the cascade  can be computed as
\begin{equation}\label{eq:csol}
  A(\omega_{\pm n})\ =\ \pm (\sqrt{p}-1) \int_{\omega_0}^{\omega_{+n}}\frac{\ud \omega_n}{\omega_n k_n}\ +\ C_{\pm}(\omega_0, A_0, p)
\end{equation}
Accordingly, by definition the energy $E_n(\omega_n) \propto A^2(\omega_n)$, which provides us with the discrete set of energies of individual harmonics. The spectral density $\E^{(\mathrm{Dir})}(\omega)$ can be now computed
\begin{equation*}
  \E^{(\mathrm{Dir})}(\omega) \doteq \lim_{\Delta\omega_n\to 0}\frac{E(\omega_{n+1}) - E(\omega_{n})}{\Delta\omega_n}.
\end{equation*}
A similar formula can be written for the inverse cascade as well. However, the limits of integration in \eqref{eq:csol} will be inverted correspondingly to $\int_{\omega_{-n}}^{\omega_{0}}$.

Besides the form of energy spectra, the $D$-cascade model allows to make other predictions for the \acf{nls}-family of the PDEs \eg.  the time scale for the $D$-cascade occurring is $t_{\mathrm{MI}}\ \propto\ t/\eps^2$; the distance between two cascading frequencies depends on the steepness of the initial wave train; the cascade can be determined, depending on the choice of excitation parameters $\omega_0, k_0, A_0$; the $D$-cascade termination can be caused by a few main reasons: stabilization, wave breaking and intermittency. All scenarios  were observed experimentally, \eg in \cite{Denissenko2007, Tulin1999}.

Connection between the \acf{nls} and  the \acs{mkdv}  can be established in the following way. As it was shown by R.~\textsc{Grimshaw} (2002) \cite{Grimshaw2002}, the following \acf{gkdv} equation
\begin{equation*}
  \mathrm{gKdV:} v_t\ +\ vv_x\ +\ v^2v_x\ +\ v_{xxx}\ =\ 0
\end{equation*}
can be regarded as weakly nonlinear model for internal long waves in a stratified fluid. This model equation can be reduced to
the \acs{mkdv} equation \eqref{mKdV} with positive nonlinearity:
\begin{equation*}
  u_t\ +\ u_{xxx}\ +\ 6u^2u_x\ =\ 0,
\end{equation*}
which in turn the classical \acf{nls} equation to an asymptotic accuracy $\O(\eps^4)$:
\begin{equation}\label{eq:nls}
  \ui\pd{A}{\tau}\ =\ 3k\pd{^2A}{\xi^2}\ +\ 6k|A|^2A,
\end{equation}
after suitable change of variables.

By derivation equation \eqref{eq:nls} describes the envelope propagation for the \acs{mkdv} solutions. Since both models are modulationally unstable, the $D$-cascade model can be applied to them.

\begin{figure}
  \centering
  \includegraphics[width=0.99\textwidth]{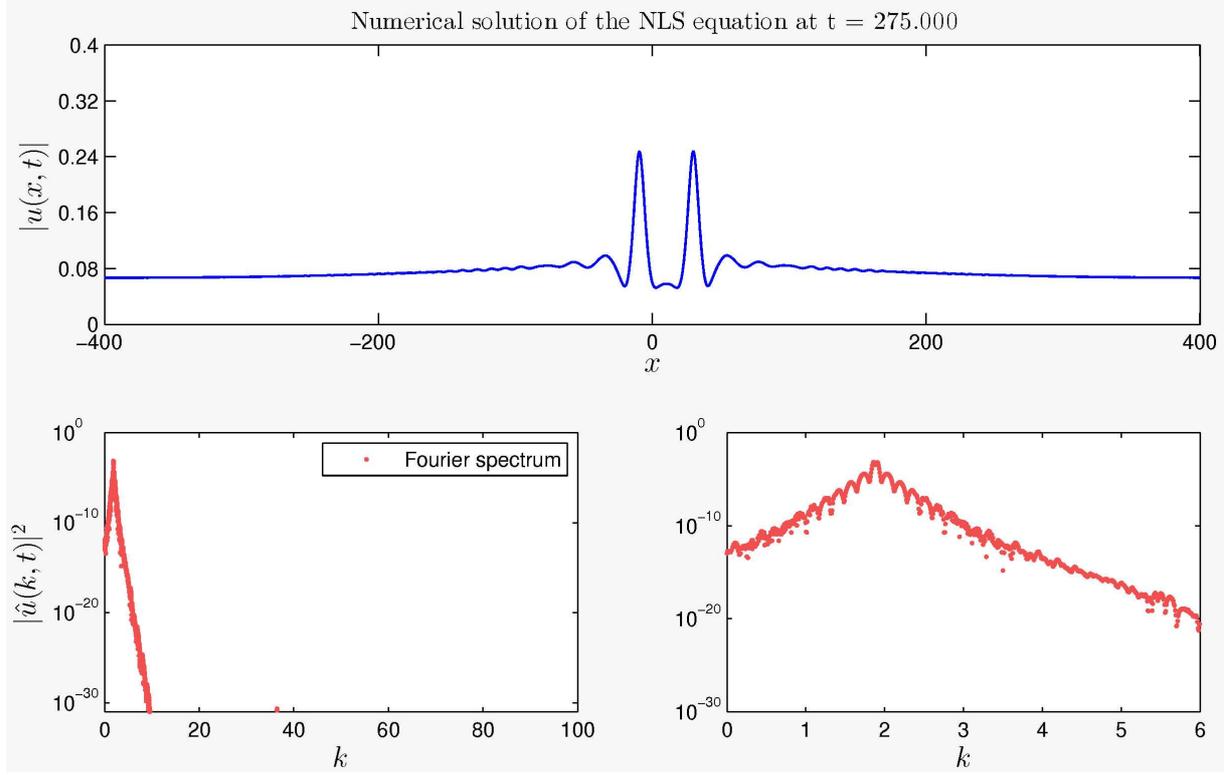}
  \caption{\small\em \acs{mi} in the NLS equation \eqref{eq:nls} for the initial condition parameters given in Table~\ref{tab:test1}.}
  \label{fig:minls}
\end{figure}

However, the \acs{nls} equation is built upon the additional assumption of the narrow band spectrum. Consequently, the $D$-cascade for a single \acs{nls} equation consists of only one cascading mode, possibly accompanied with the spectrum broadening. This situation is illustrated on Figure~\ref{fig:minls} where equation \eqref{eq:nls} was solved numerically for a modulated plane wave initial condition.

The full description of a non-trivial cascade in the \acs{nls} framework requires a sequence of the \acs{nls} models: each \acs{nls} equation describing the vicinity of a cascading mode. On the other hand, the \acs{mkdv} equation does not have such restrictions. Consequently, we can expect to observe the full $D$-cascade in this model. Numerical study  of the inverse $D$-cascade in the \acs{mkdv} equation is given in \cite{Dutykh2014b}, while below we focus on the direct $D$-cascade formation in the same equation.

%%%%%%%%%%%%%%%%%%%%%%%%%%%%%%%%%%%%%%%%%%%%%%%%%%%%%%%%%%%%%%%%%%%%%%

\section{Numerical simulations}\label{s:numerics}

In order to solve numerically the \acs{kdv} equation on a periodic domain we used the classical Fourier-type pseudo-spectral method \cite{Boyd2000, Trefethen2000}. The derivatives are computed in the Fourier space, while the nonlinear products -- in the physical one (with the linear CPU-time). Thanks to the FFT algorithm \cite{FFTW98, FFTWgen99, Frigo2005} the passage between these two representations is done in the super-linear time $\O(N\log(N))$, which determines the overall algorithm complexity (per time step). For the dealiasing we used the classical $2/3$-rule \cite{Trefethen2000} which was combined (when necessary) with the Fourier smoothing method proposed in \cite{Hou2007} for more delicate treatment of higher frequencies. The discretization in time was done with the embedded adaptive 5\up{th} order Cash--Karp Runge--Kutta scheme \cite{Cash1990} with the adaptive PI step size control \cite{Hairer1996}.

As there are several exact analytical solutions known for the \eqref{mKdV+} equation, we use them for the verifying our numerical model. The exact cnoidal wave solutions to the \acs{kdv} equation can be found in \cite{Bronski2011}:
\begin{equation}\label{eq:cnoidal}
  u(x,t) = \frac{1}{\frac{s}{3s^2-1} + \frac{\sqrt{2(1 - s^2)}}{6s^2-2}\sin\bigl(\sqrt{3s^2 - 1}(x - x_0 - t)\bigr)} - s, \quad s\in \Bigl[-1, -\frac{\sqrt{3}}{3}\Bigr] \cup \Bigl[\frac{\sqrt{3}}{3}, 1\Bigr], \quad x_0\in\R.
\end{equation}
The one-soliton solution is given by
\begin{equation}\label{eq:1soliton}
  u(x,t) = a + \frac{b^2}{\sqrt{4a^2+b^2} \cosh y + 2a}, \quad y = b x - (6a^2b + b^3)t + c\ ,
\end{equation}
where $a$, $b$, $c$ are arbitrary constants, \cite{Miura1968}. There exists also a rational solution of the form
\begin{equation*}%\label{rational}
  u(x,t) = a - \frac{4a}{4a^2(x - 6a^2t)^2 + 1}\ ,
\end{equation*}
where $a$ is arbitrary constant, \cite{Ono1976}.

The numerical solver was validated first on the simple tests of the cnoidal wave propagation  and the overtaking collision of solitary waves. We checked that up to the numerical accuracy the spectrum was stationary in the former simulation showing that the wave is perfectly preserved by the numerical solver and no dispersive tail was present after the collision in the latter, confirming the integrability of the \acs{mkdv} equation.

For our numerical simulations of the \acf{mi} we adopt the set-up used also earlier in \cite{Grimshaw2005}. We consider the following initial condition posed on a periodic domain $[-\ell, \ell] = \bigl[-{\pi}/{k_0}, {\pi}/{k_0}\bigr]$:
\begin{equation}\label{eq:pert}
  u(x,0) \equiv u_0(x) = a\bigl(1 + \delta\sin(K_0 x)\bigr)\sin(k_0 x),
\end{equation}
where $a$ is the base wave amplitude, $\delta$ is the perturbation magnitude and the wavenumbers $k_0$, $K_0$ are chosen such that their ratio $k_0/K_0\in\Z$. The values of parameters are given in the Table~\ref{tab:test1}. The number $N$ of Fourier modes used in simulations presented below will be set to $N = 32768$ and the tolerance parameter in the PI step size control was set to $10^{-8}$.

\begin{table}
  \centering
  \begin{tabular}{|>{\columncolor[gray]{0.85}}l||>{\columncolor[gray]{0.85}}c|}
  \hline\hline
    Base wave amplitude, $a$ & $0.08$ \\
    Perturbation magnitude, $\delta$ & $0.05$ \\
    Base wavenumber, $k_0$ & $1.884$ \\
    Perturbation wavenumber, $K_0$ & $0.00785$ \\
    Ratio of wavelengths, $k_0/K_0$ & $240$ \\
  \hline\hline
  \end{tabular}
  \bigskip
  \caption{\small\em Numerical parameters from the previous study \cite{Grimshaw2005} used for the comparison with the present results.}
  \label{tab:test1}
\end{table}

\subsection{Effect of the spectral domain}

For the Fourier domain limited to $k\in[0,\ 8]$ the simulation results shown on Figures~\ref{fig:MI1} and \ref{fig:MI2} are in a good qualitative agreement with the results presented in \cite{Grimshaw2005}. Unfortunately the authors in \cite{Grimshaw2005} did not report the evolution of the Fourier spectrum. We will fill in this gap in the present study. For instance, one can notice the presence of a second peak in the Fourier spectrum (in the vicinity of $k\ \approx\ 6$) which corresponds to the second cascading mode. During the development of the \acs{mi} we observe a broadening of the spectrum around two cascading modes present in this numerical simulation. This broadening continues until the \acs{mi} is fully developed (see Figure~\ref{fig:MI2}(\textit{c, d})). The left and right broadening wings can appear to be symmetric up to the graphical resolution, however our measurements reported in Table~\ref{tab:ampl} show that it is not actually the case. The broadening intensities defined as the ratio of energies of two consecutive modes, \ie $\beta_{\mathrm{r,l}}\ \doteq\ \frac{|u_{k\pm 1}|^2}{|u_k|^2}$ and  observed in the experiments depend on the initial wave amplitude. Namely, when the nonlinearity is increased, the broadening tails become flatter, thus involving more Fourier harmonics into this process.

\begin{figure}
  \centering
  \subfigure[Solution at $t = 100.0$]%
  {\includegraphics[width=0.48\textwidth]{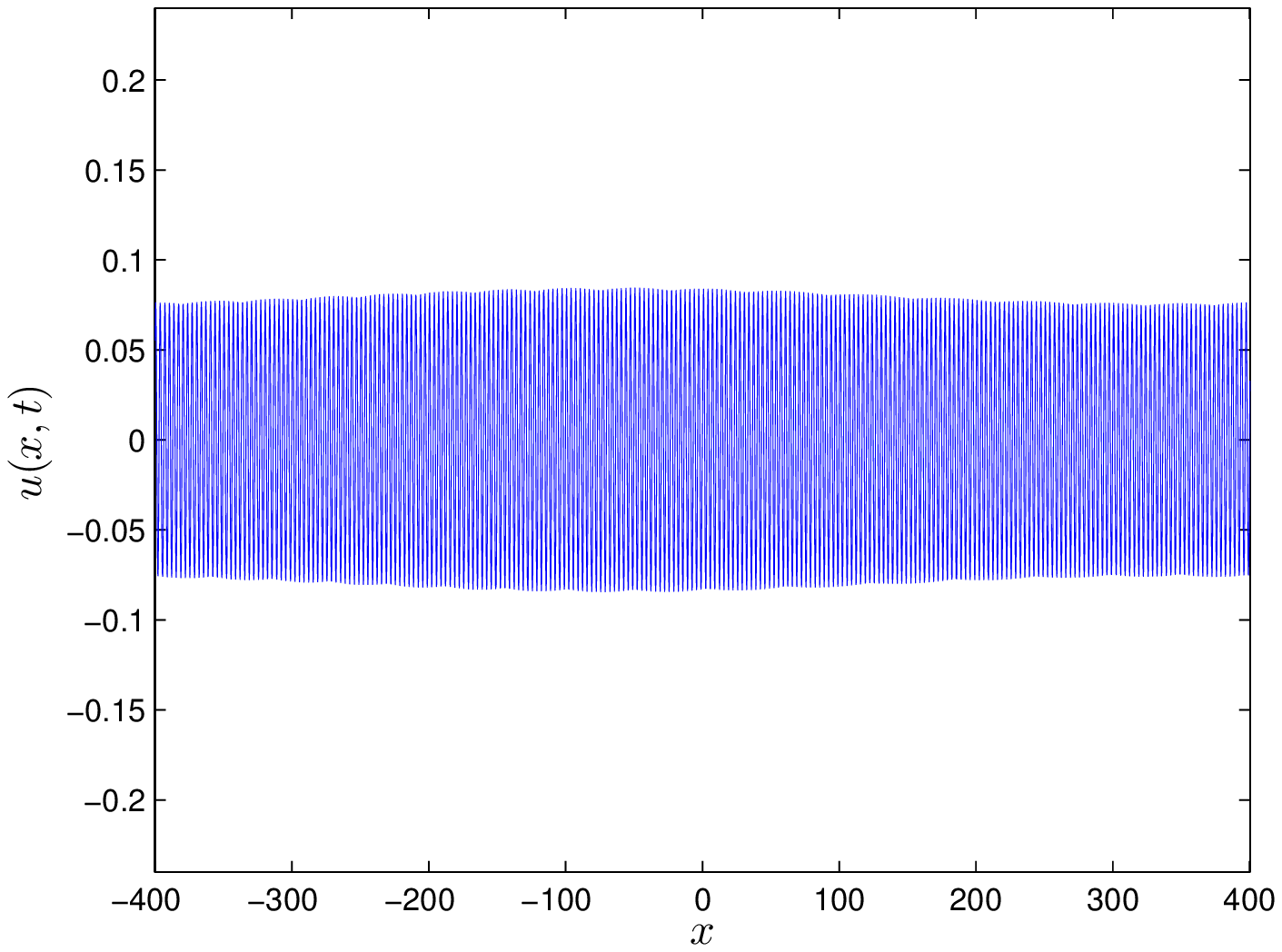}}
  \subfigure[Fourier spectrum at $t = 100.0$]%
  {\includegraphics[width=0.48\textwidth]{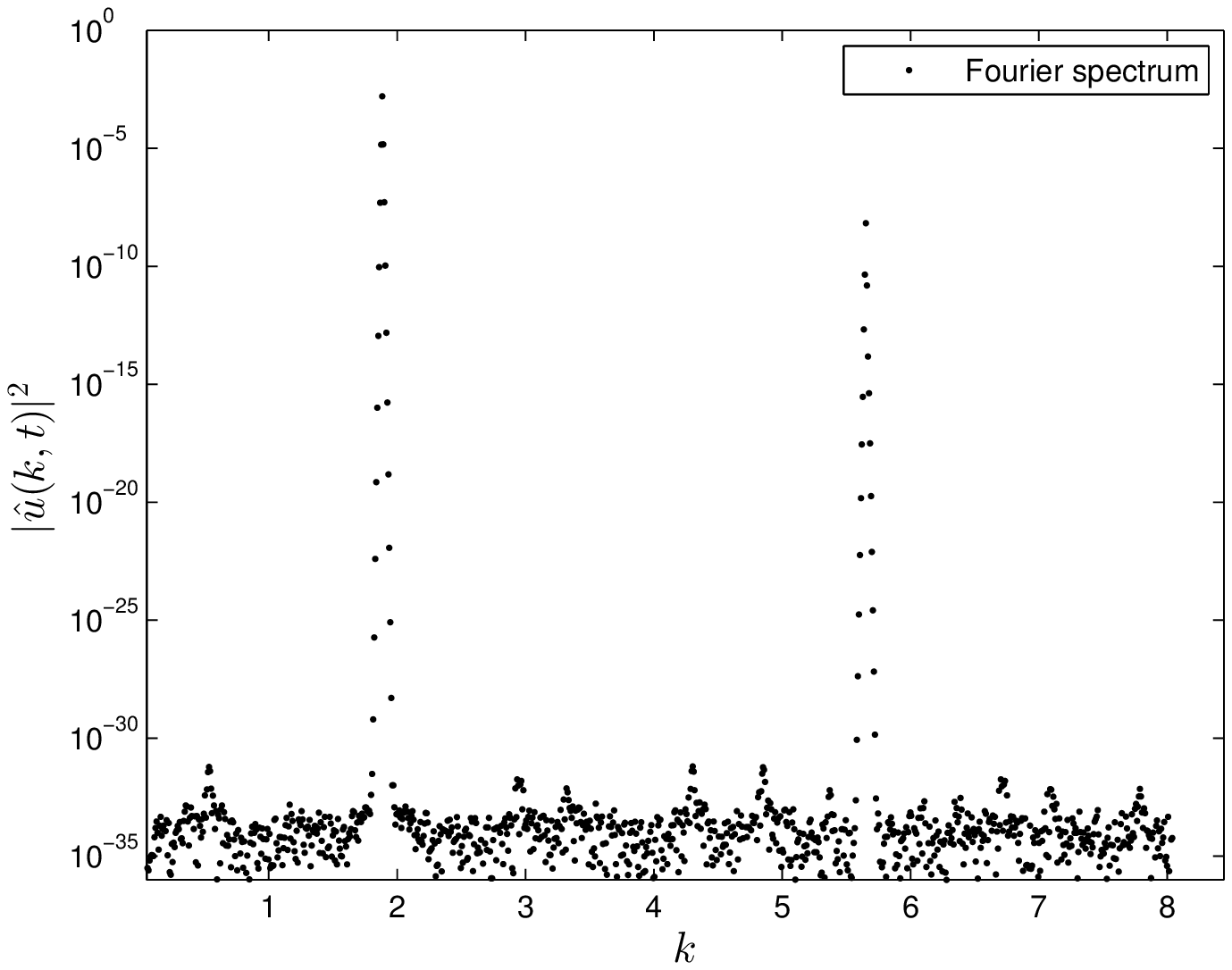}}
  \subfigure[Solution at $t = 400.0$]%
  {\includegraphics[width=0.48\textwidth]{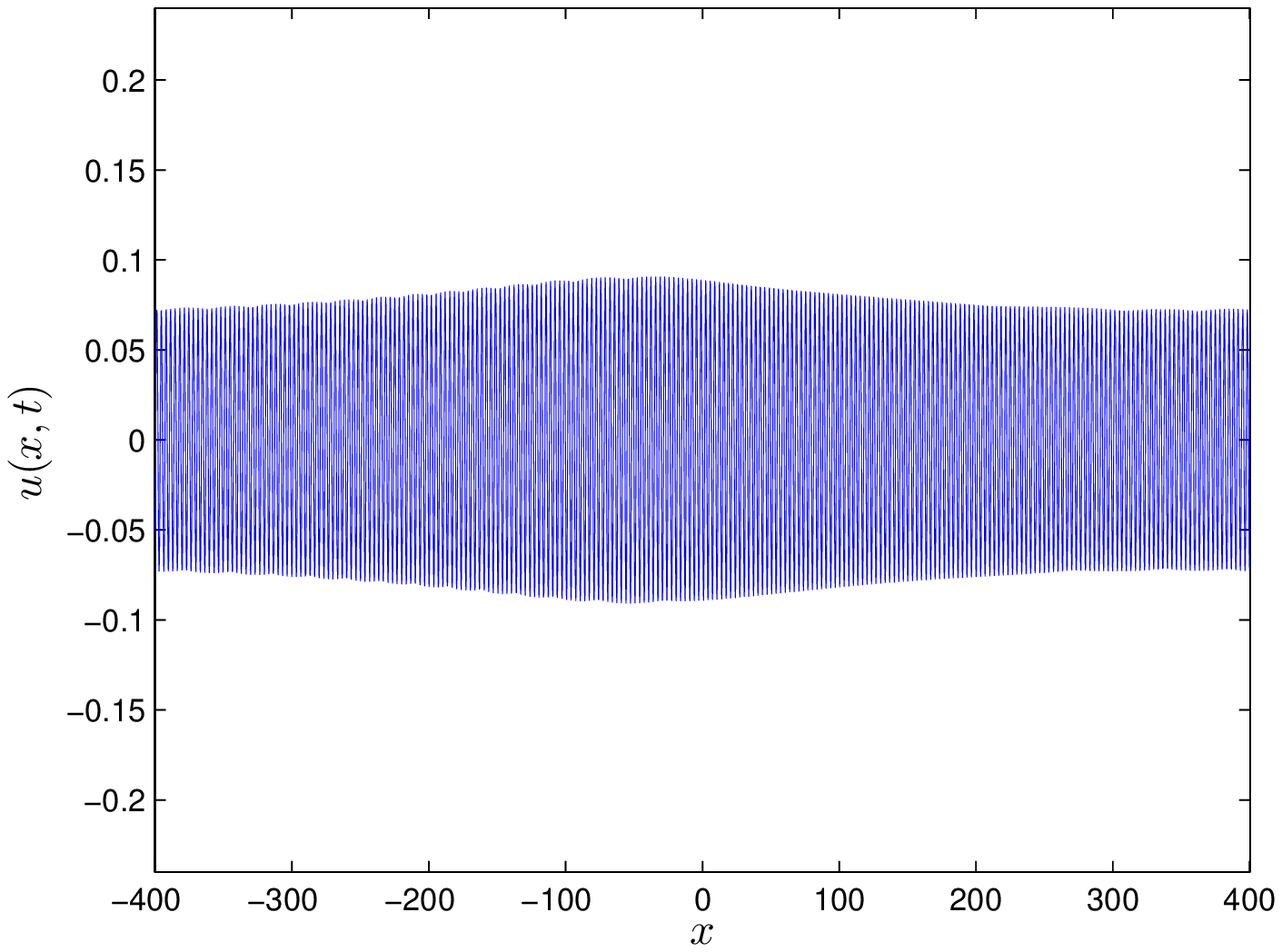}}
  \subfigure[Fourier spectrum at $t = 400.0$]%
  {\includegraphics[width=0.48\textwidth]{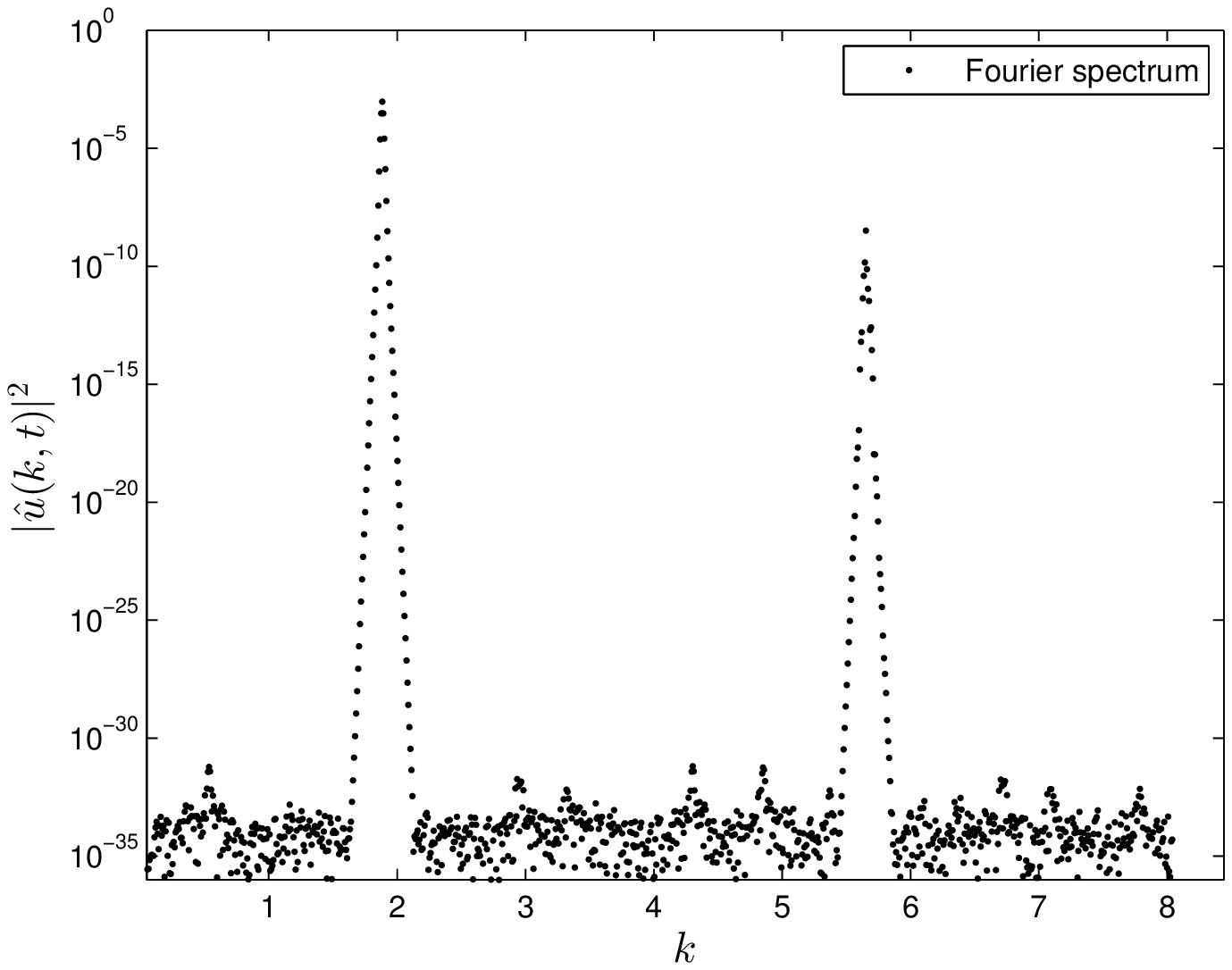}}
  \subfigure[Solution at $t = 510.0$]%
  {\includegraphics[width=0.48\textwidth]{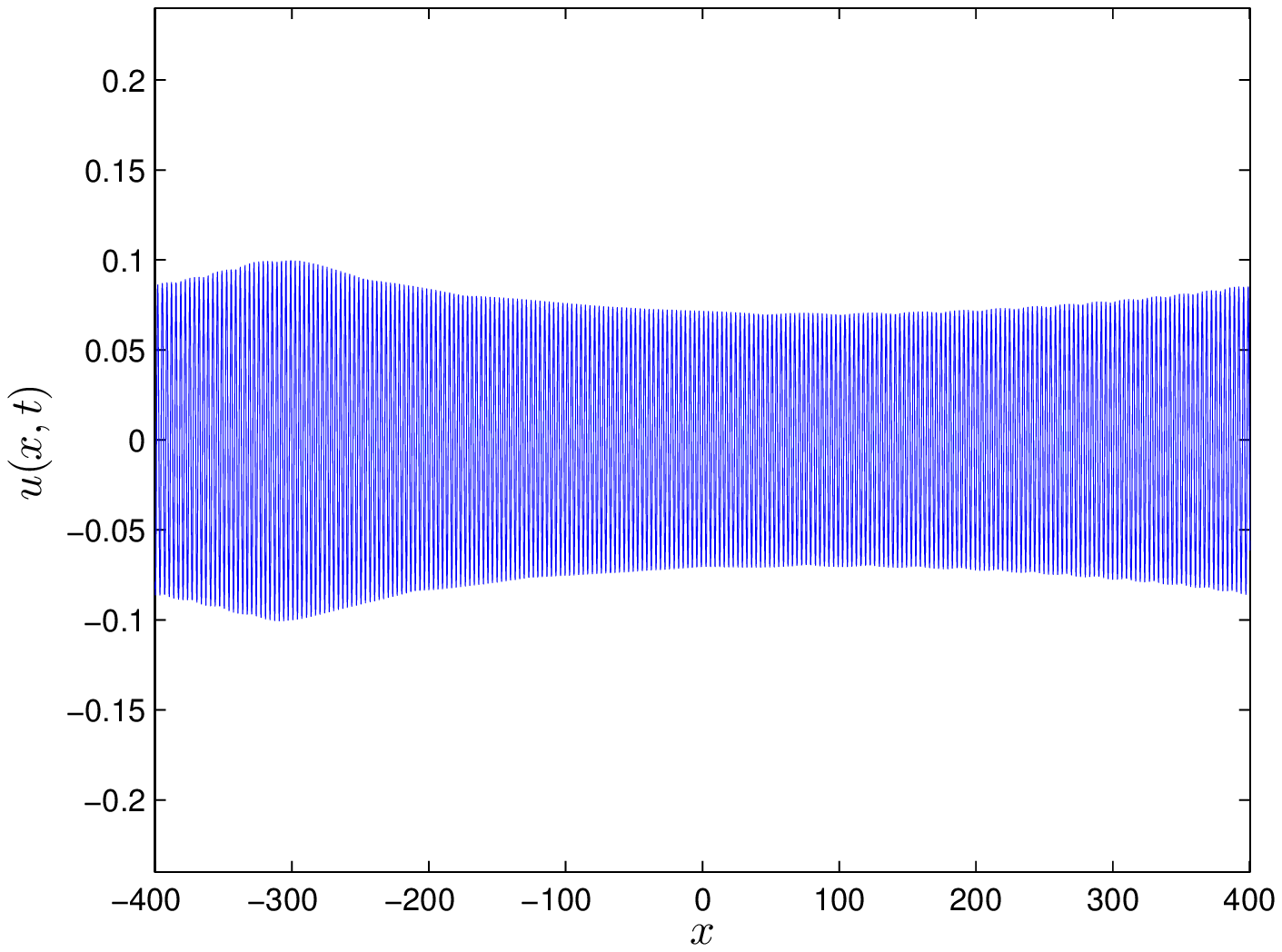}}
  \subfigure[Fourier spectrum at $t = 510.0$]%
  {\includegraphics[width=0.48\textwidth]{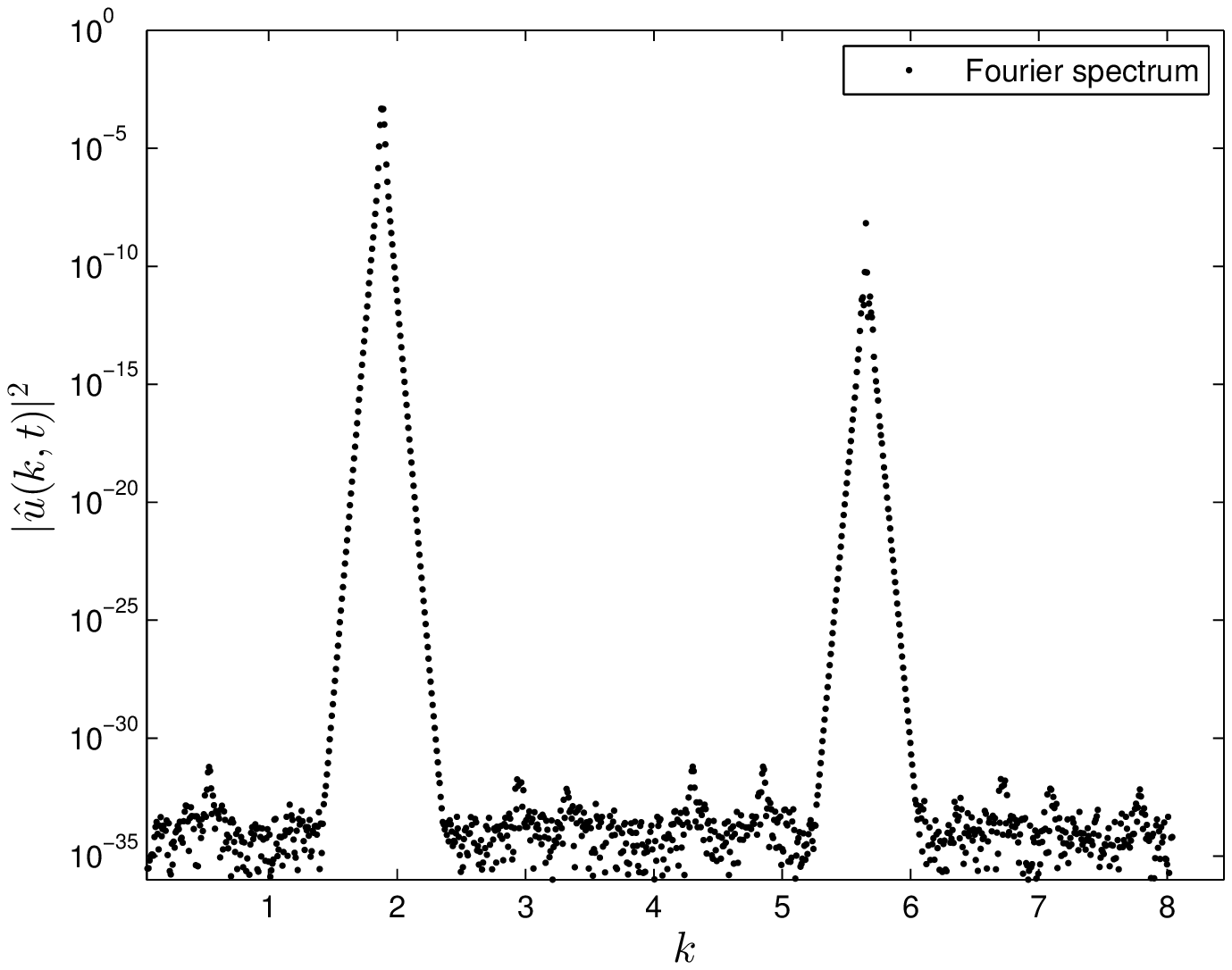}}
  \caption{\small\em Development of the modulational instability in the \acs{mkdv} equation for parameters given in Table~\ref{tab:test1}. The left panel shows the \acs{mkdv} solution and on the right panel we show the Fourier spectrum for $k\in[0,\ 8]$.}
  \label{fig:MI1}
\end{figure}

\begin{figure}
  \centering
  \subfigure[Solution at $t = 560.0$]%
  {\includegraphics[width=0.48\textwidth]{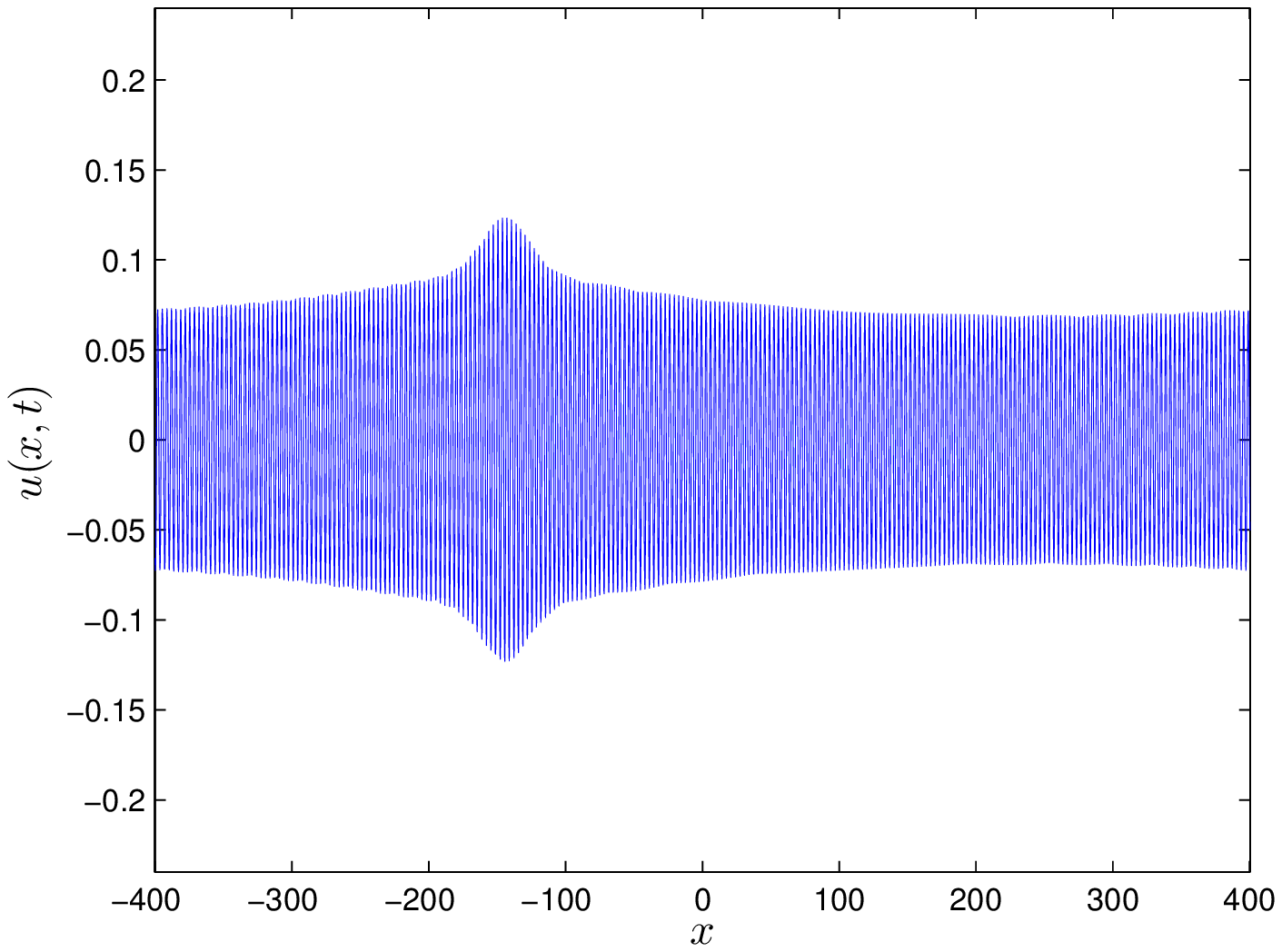}}
  \subfigure[Fourier spectrum at $t = 560.0$]%
  {\includegraphics[width=0.48\textwidth]{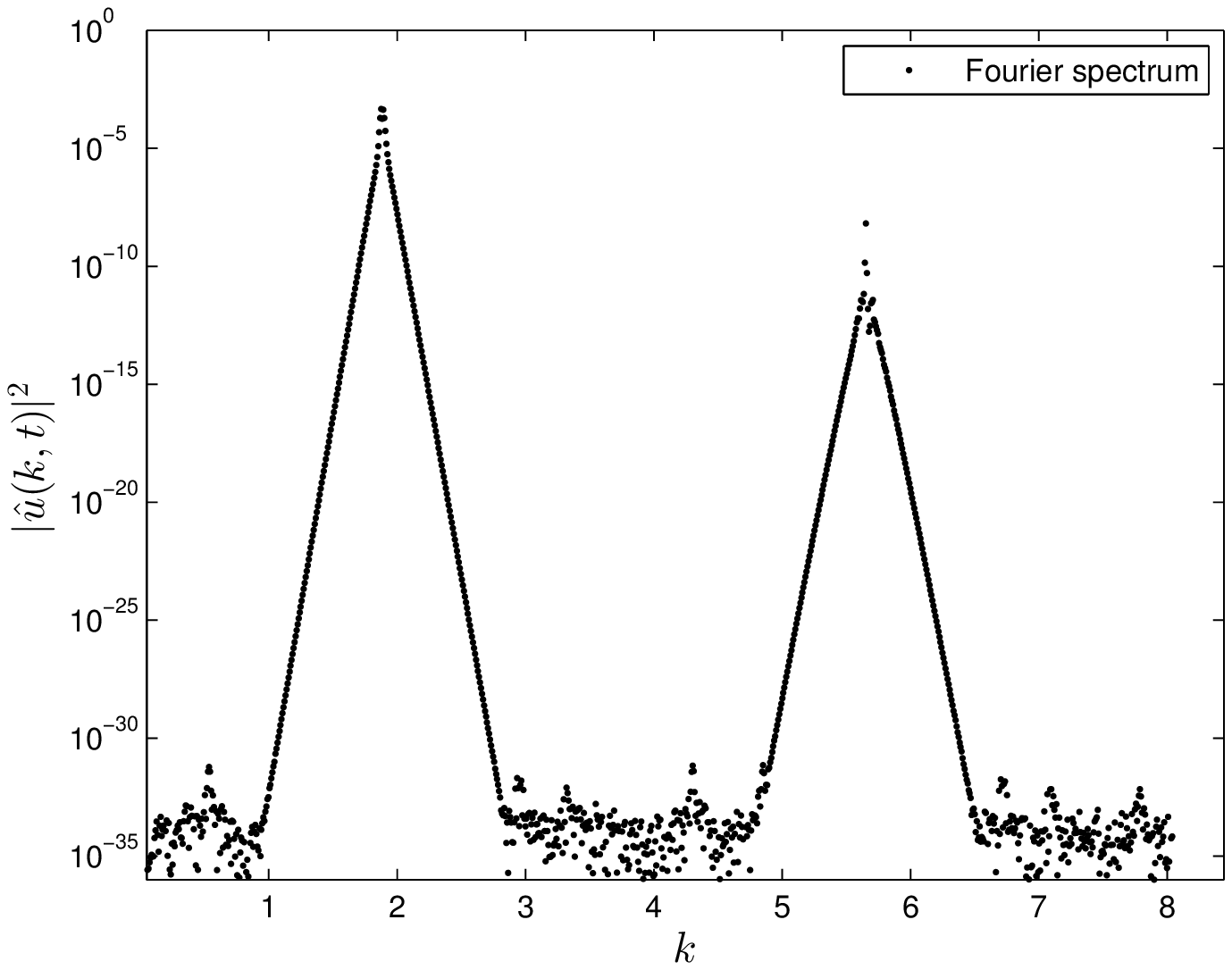}}
  \subfigure[Solution at $t = 600.0$]%
  {\includegraphics[width=0.48\textwidth]{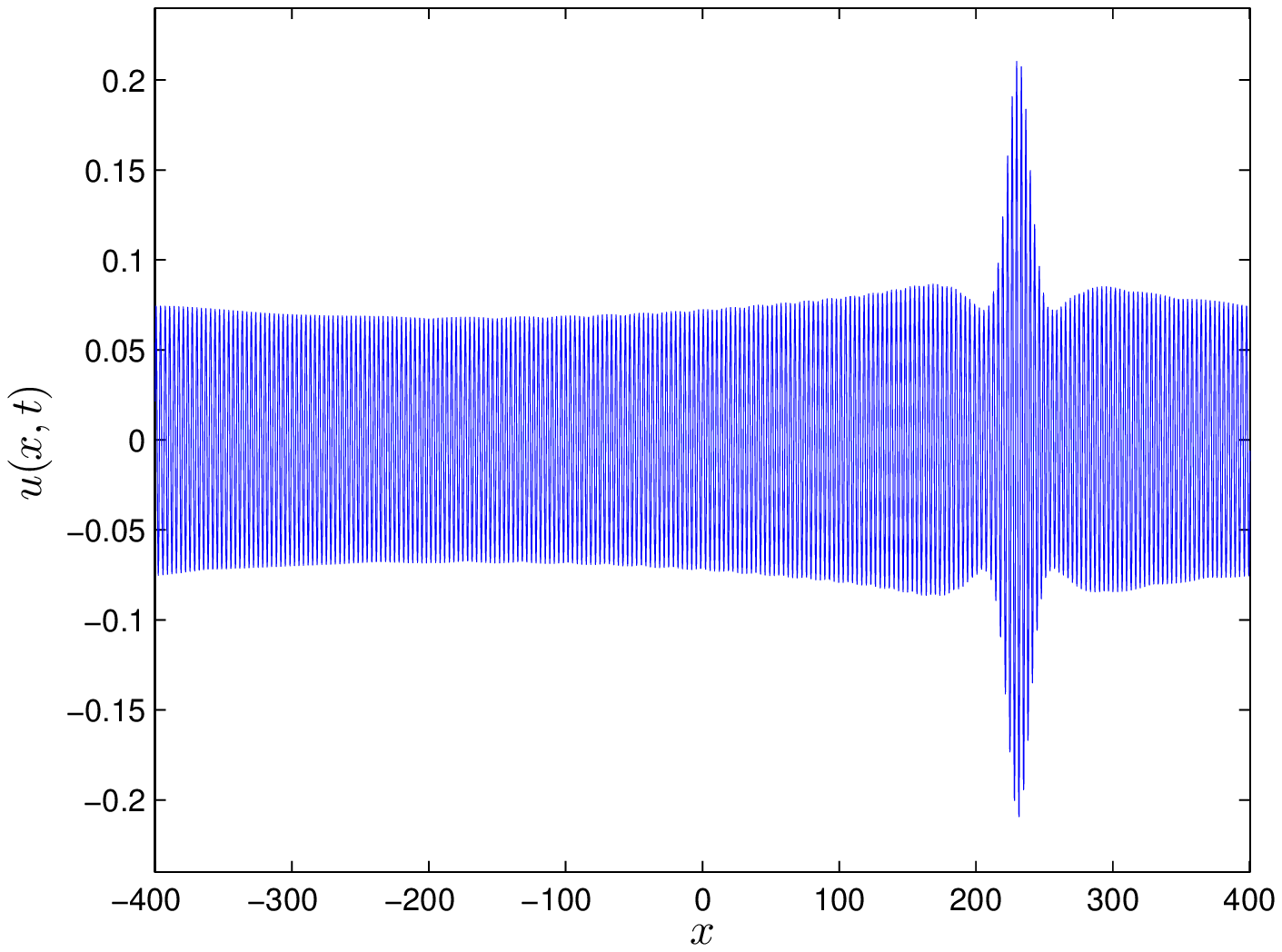}}
  \subfigure[Fourier spectrum at $t = 600.0$]%
  {\includegraphics[width=0.48\textwidth]{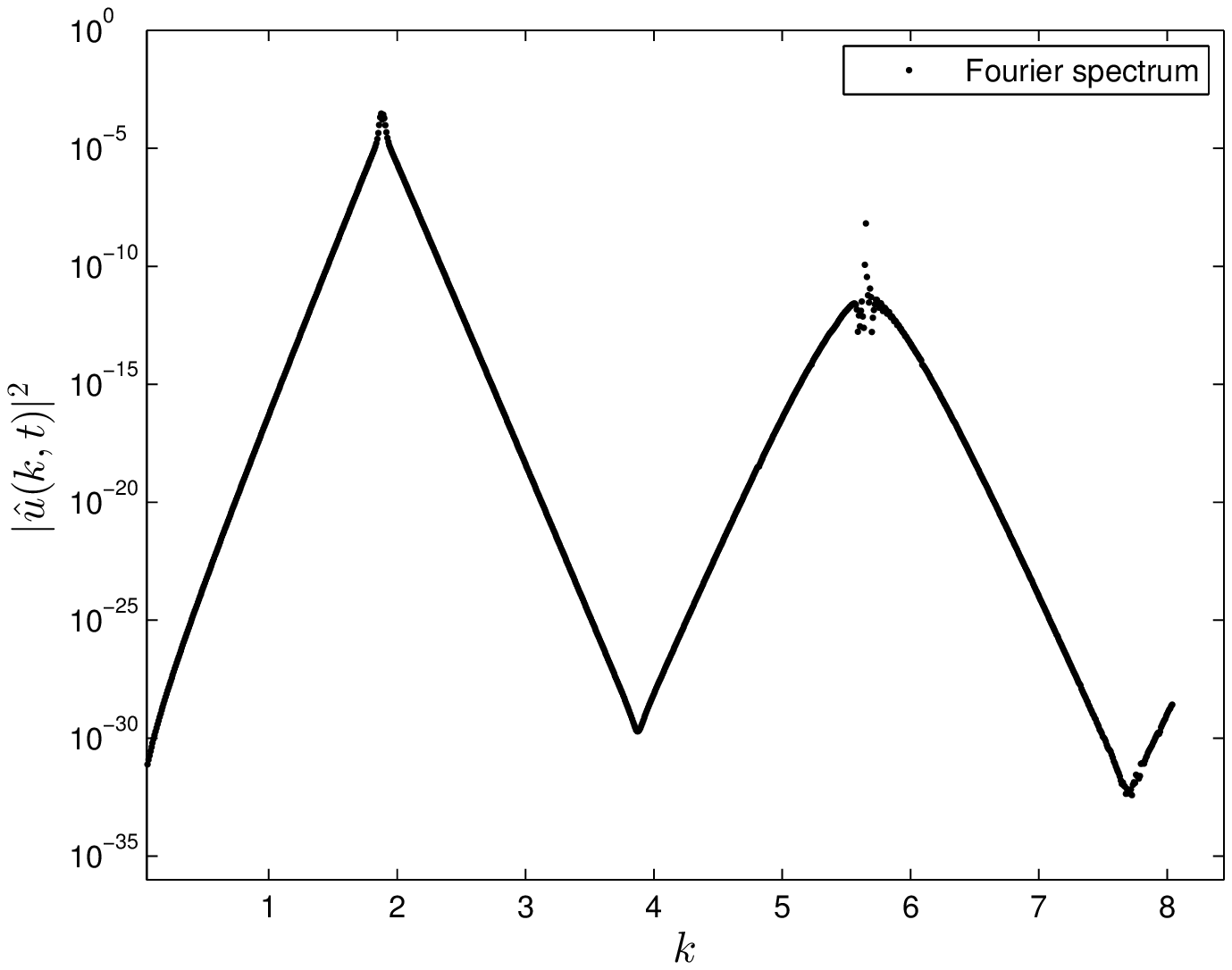}}
  \subfigure[Solution at $t = 700.0$]%
  {\includegraphics[width=0.48\textwidth]{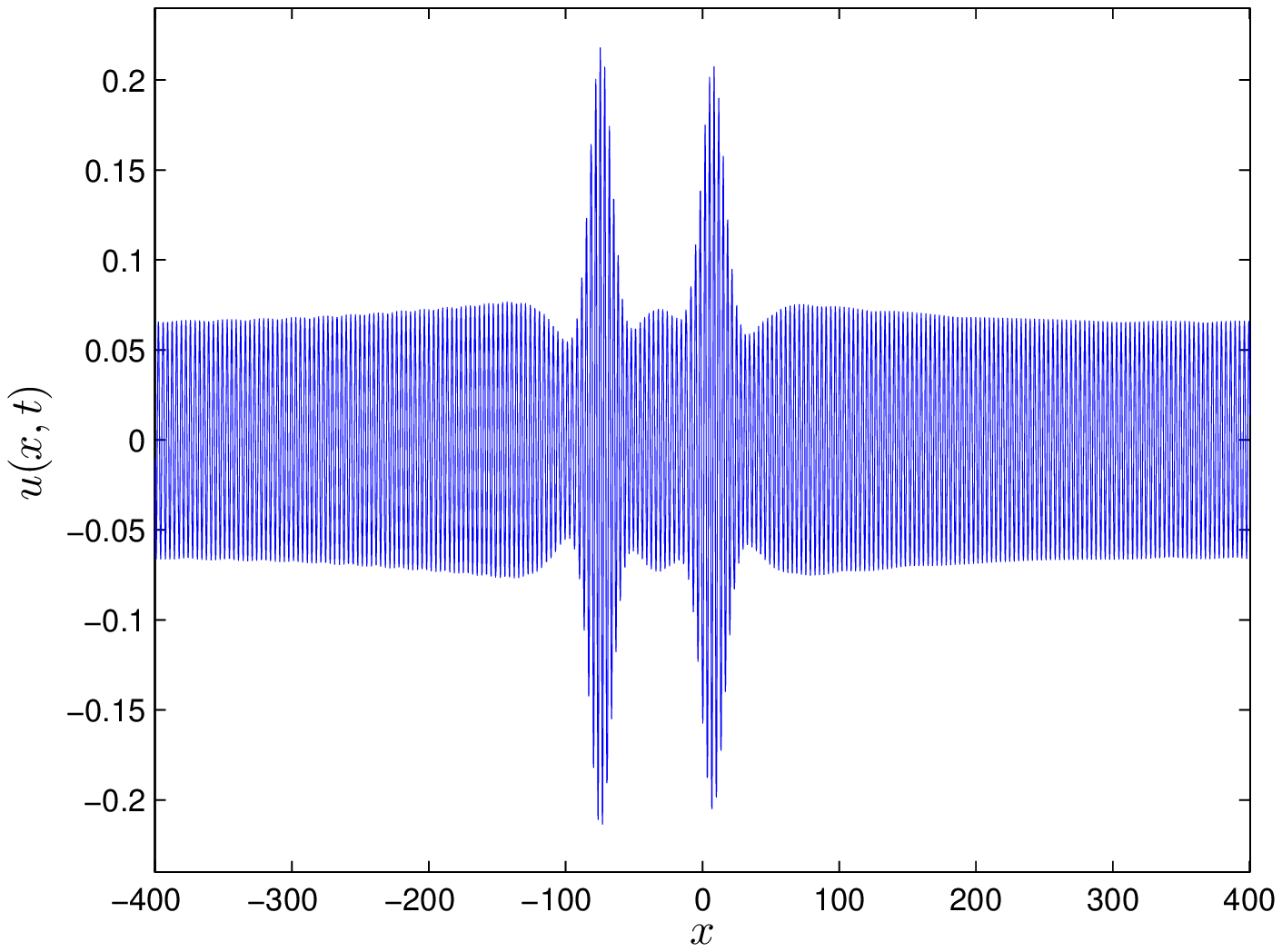}}
  \subfigure[Fourier spectrum at $t = 700.0$]%
  {\includegraphics[width=0.48\textwidth]{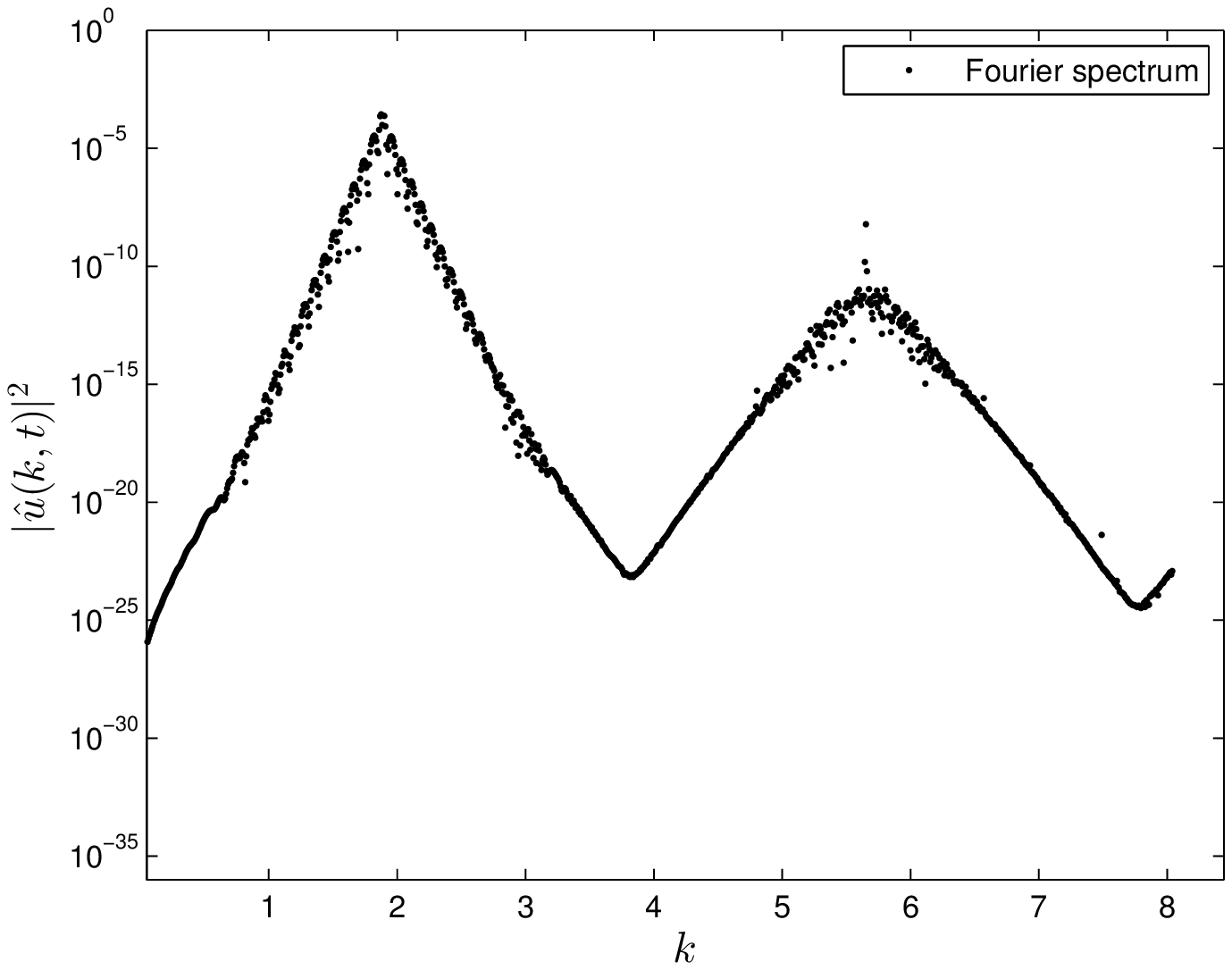}}
  \caption{\small\em (Continued). See Figure~\ref{fig:MI1} for the detailed description.}
  \label{fig:MI2}
\end{figure}

\begin{table}
  \centering
  \begin{tabular}{|>{\columncolor[gray]{0.85}}c||>{\columncolor[gray]{0.85}}c||>{\columncolor[gray]{0.85}}c|}
  \hline\hline
    \textit{Amplitude}, $a$ & \textit{Left intensity}, $\beta_l$ & \textit{Right intensity}, $\beta_r$ \\
  \hline\hline
    $0.02$ & $2.109$ & $2.085$ \\
    $0.03$ & $1.590$ & $1.550$ \\
    $0.04$ & $1.406$ & $1.367$ \\
    $0.05$ & $1.340$ & $1.314$ \\
    $0.08$ & $1.292$ & $1.253$ \\
  \hline\hline
  \end{tabular}
  \bigskip
  \caption{\small\em The left and right spectrum broadening intensities measured in numerical simulations of the \acs{mkdv} equation for different values of the base wave amplitude. The measurements were made at the fully developed \acs{mi}.}
  \label{tab:ampl}
\end{table}

In order to observe more than two cascading modes, the Fourier space has to be broaden. These results are shown on the left panel of Figure~\ref{fig:mi2}. In this case the Fourier domain is almost four times larger going up to $k\ =\ 32$. Accordingly to the theoretical predictions, the direct $D$-cascade is formed from the first instances of the numerical simulation. The linear fit in the Fourier space suggests the exponential shape of the energy spectrum $\E_k\ \propto\ \exp(-\alpha k)$, for some positive value of the slope $\alpha$. Moreover, we computed the distances between two successive cascading modes. It turns out that this value is constant to the numerical precision (obviously, for a fixed initial condition).

\subsection{Effect of the excitation amplitude}

\begin{figure}
  \centering
  \subfigure[$a\ =\ 0.08$, $t\ =\ 60$]{
  \includegraphics[width=0.48\textwidth]{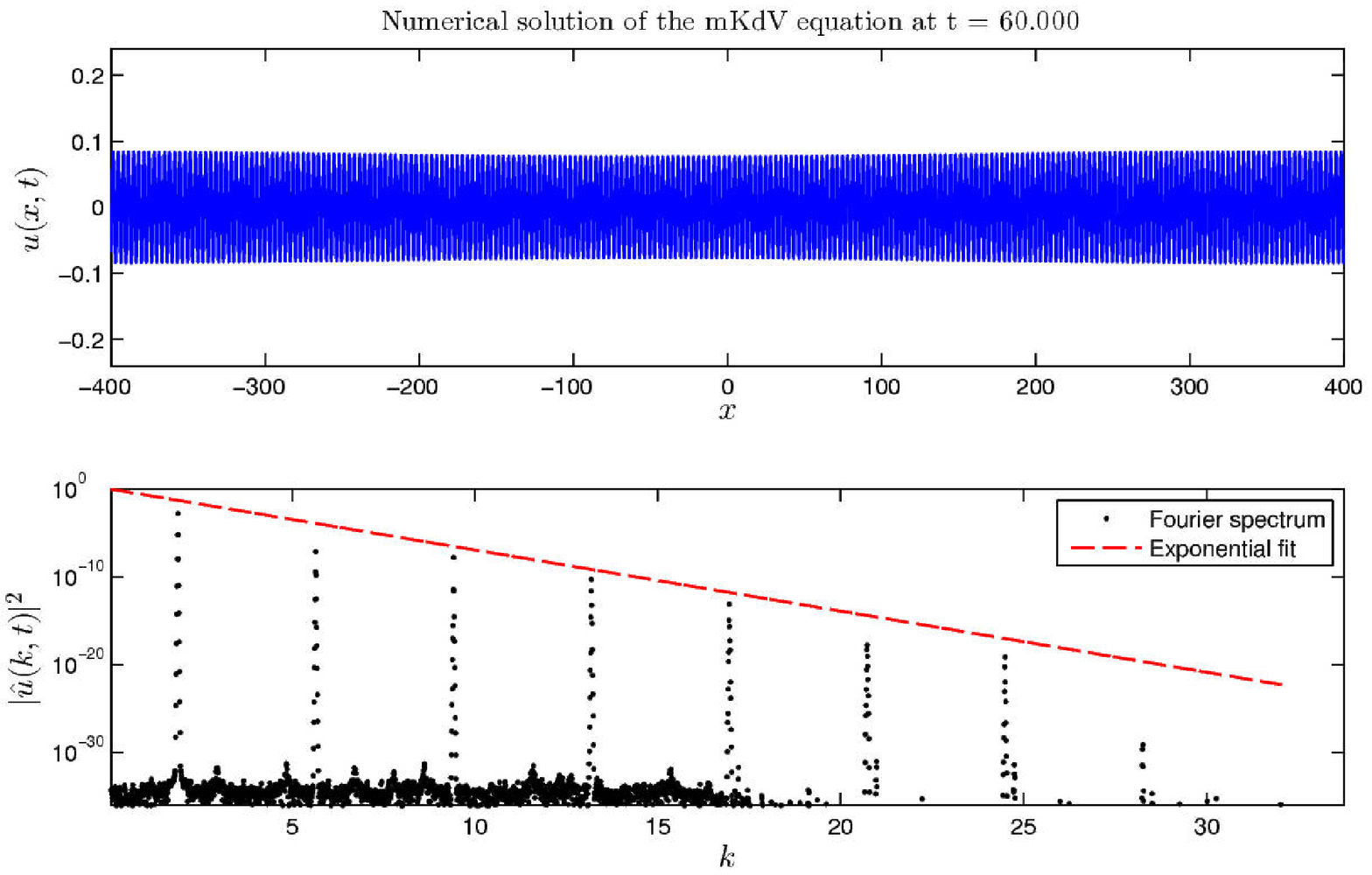}}
  \subfigure[$a\ =\ 0.16$, $t\ =\ 60$]{
  \includegraphics[width=0.48\textwidth]{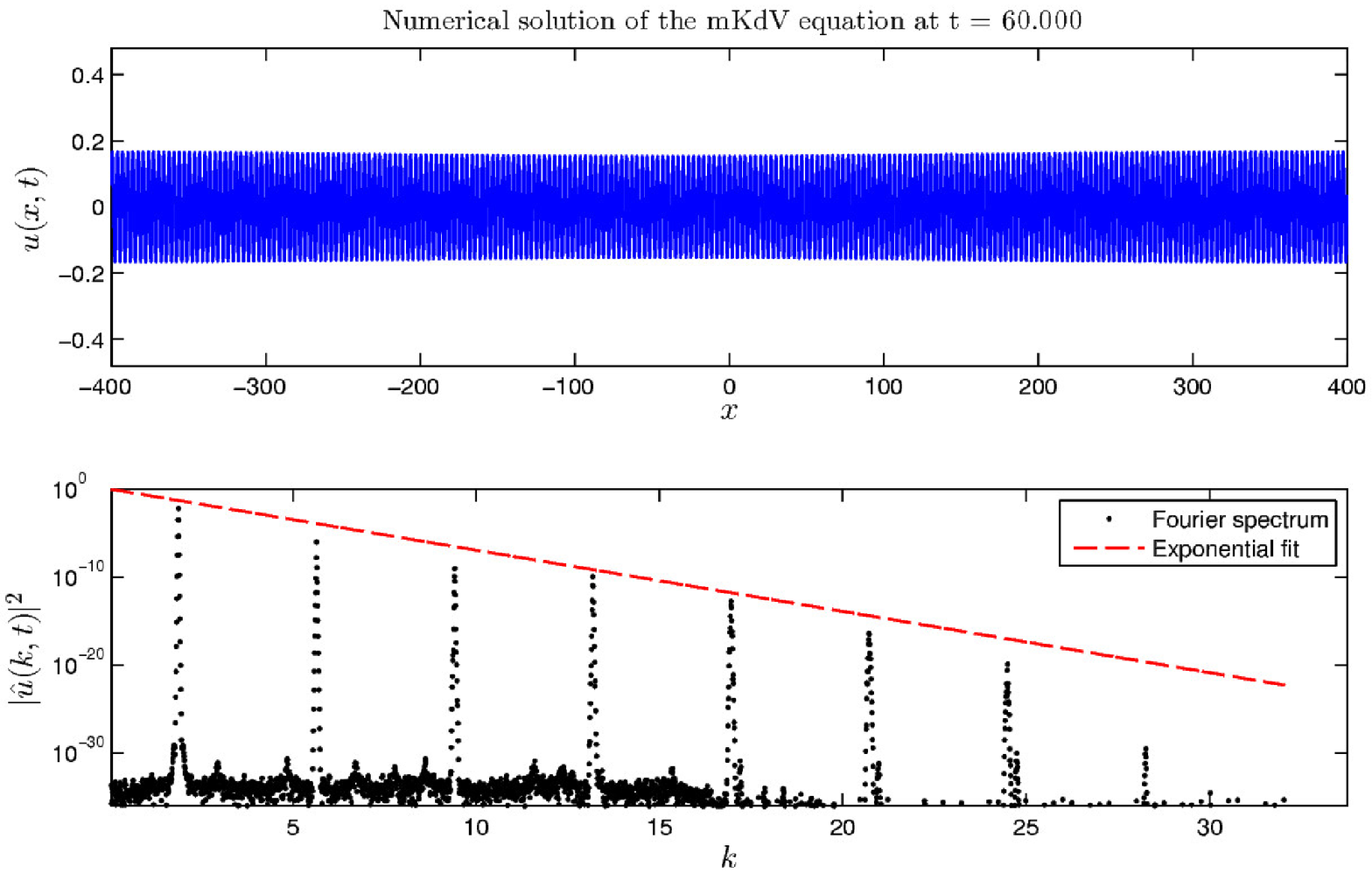}}
  \subfigure[$a\ =\ 0.08$, $t\ =\ 400$]{
  \includegraphics[width=0.48\textwidth]{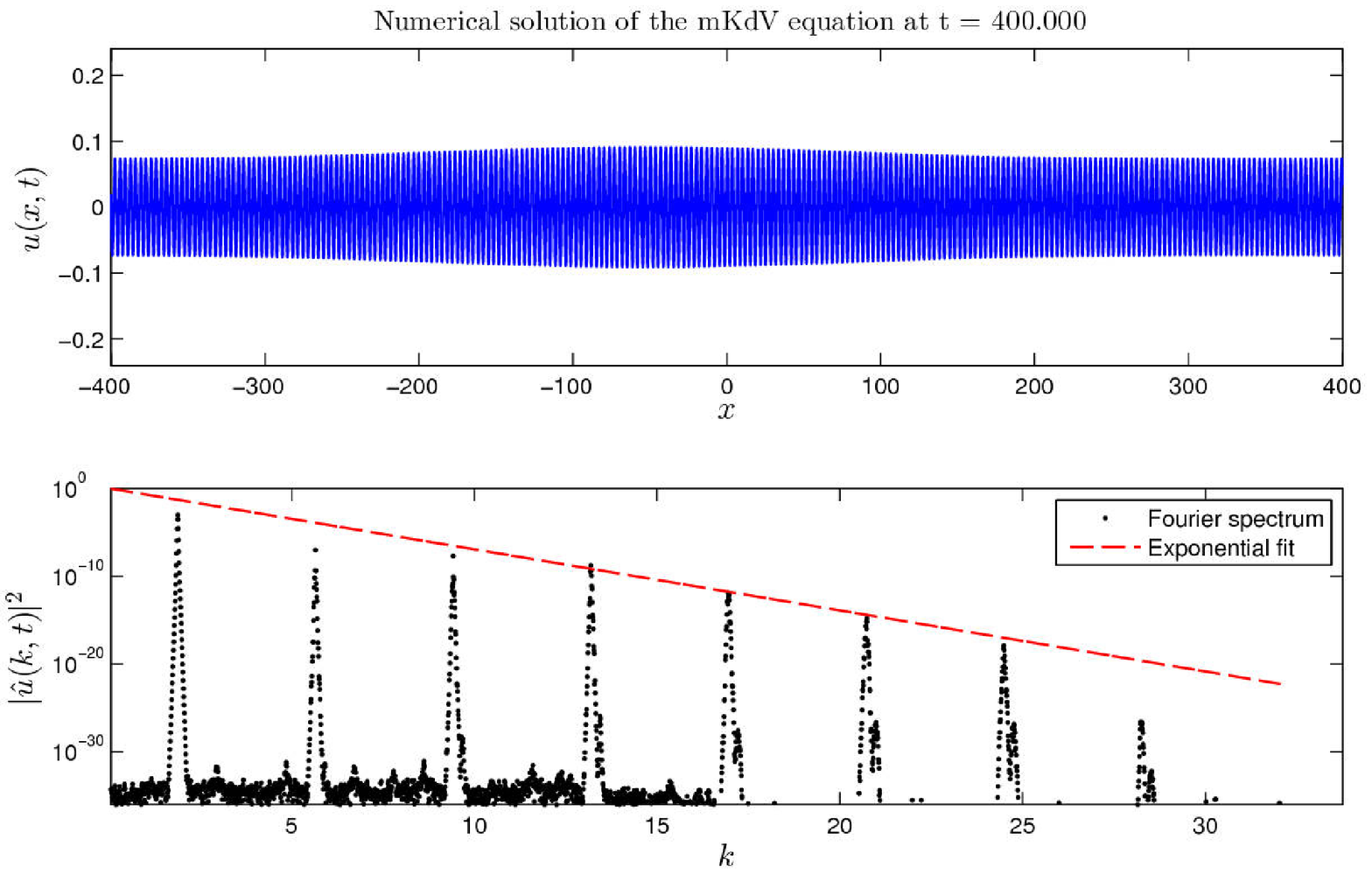}}
  \subfigure[$a\ =\ 0.16$, $t\ =\ 400$]{
  \includegraphics[width=0.48\textwidth]{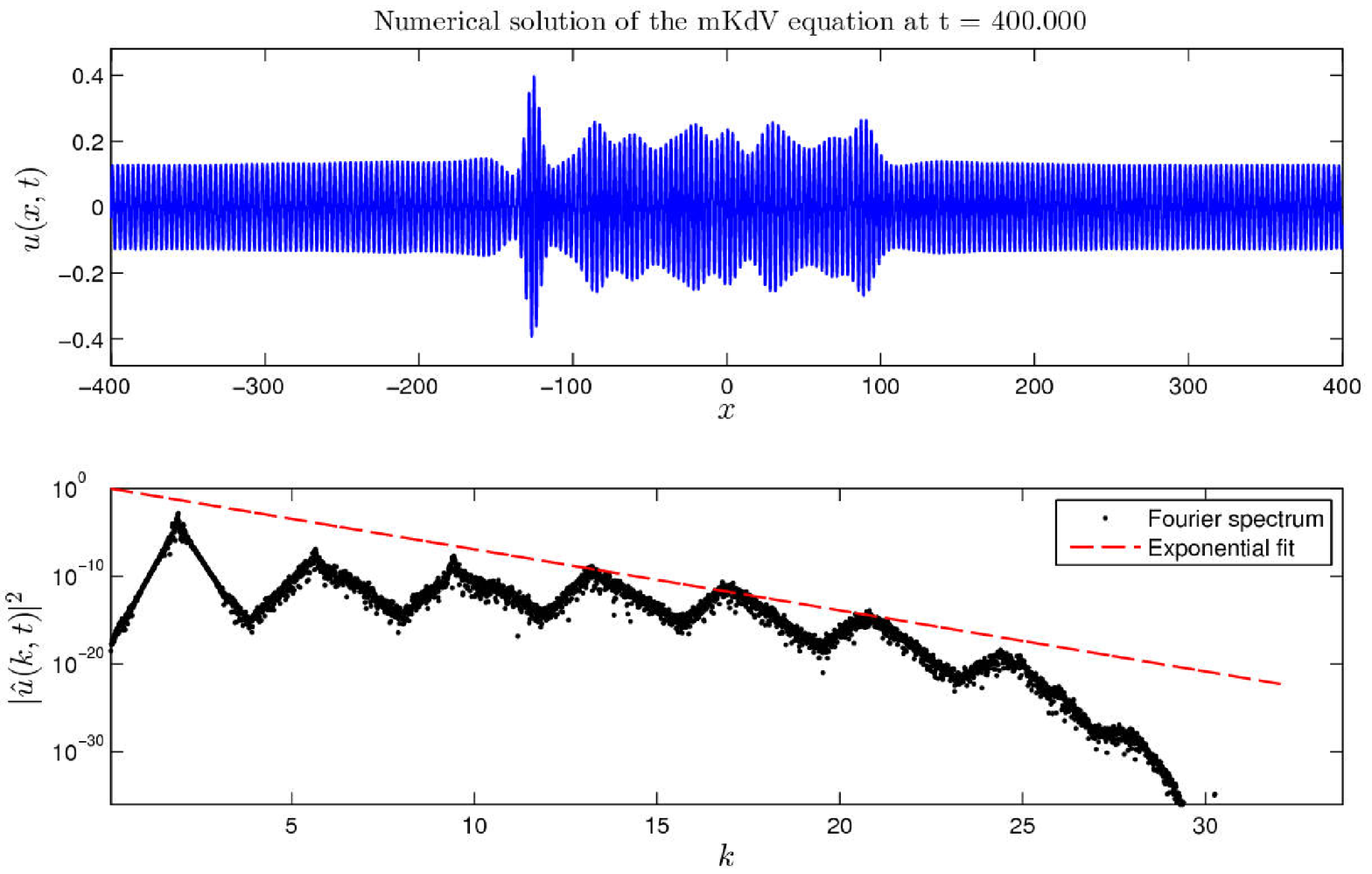}}
  \subfigure[$a\ =\ 0.08$, $t\ =\ 840$]{
  \includegraphics[width=0.48\textwidth]{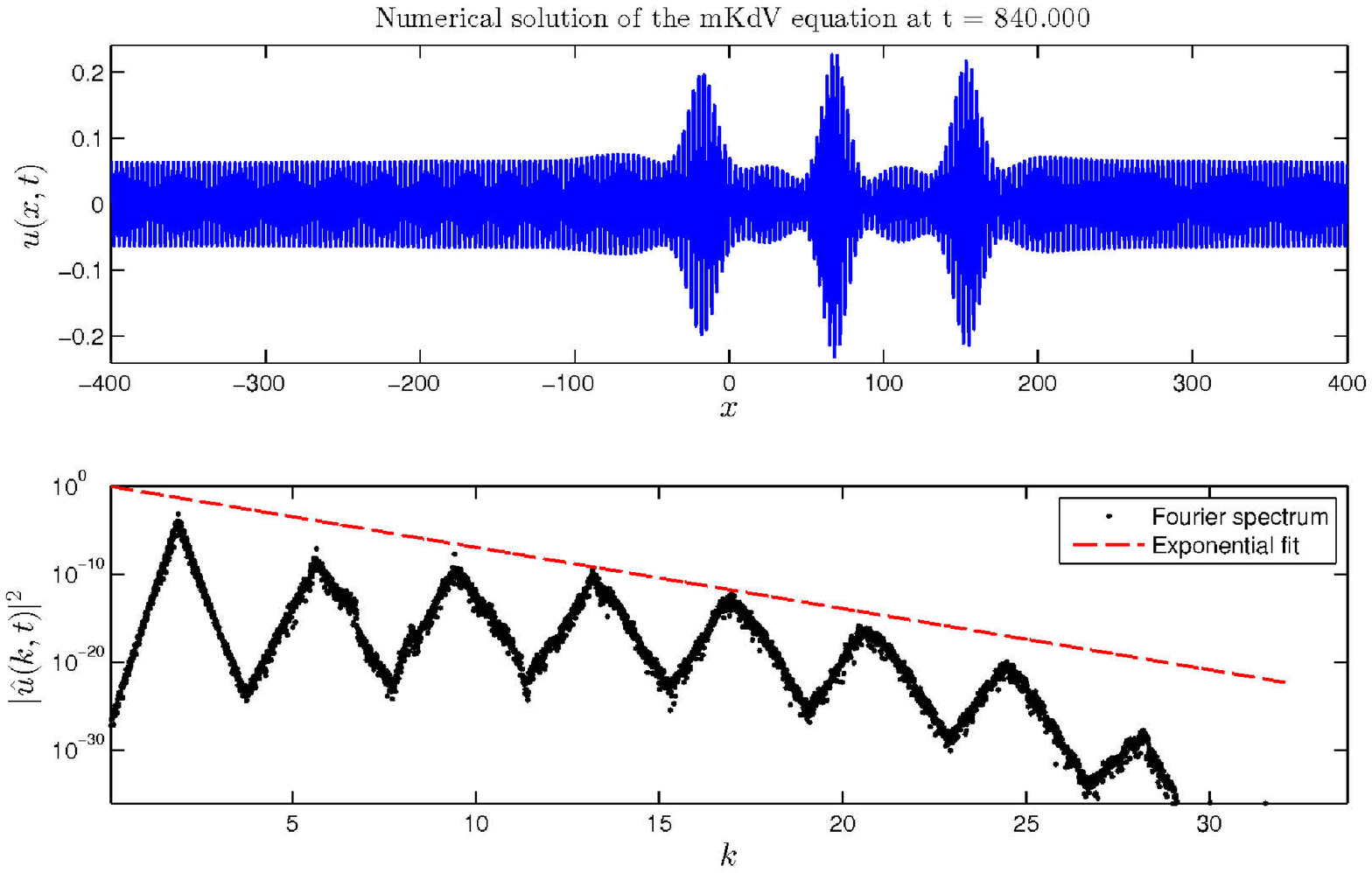}}
  \subfigure[$a\ =\ 0.16$, $t\ =\ 840$]{
  \includegraphics[width=0.48\textwidth]{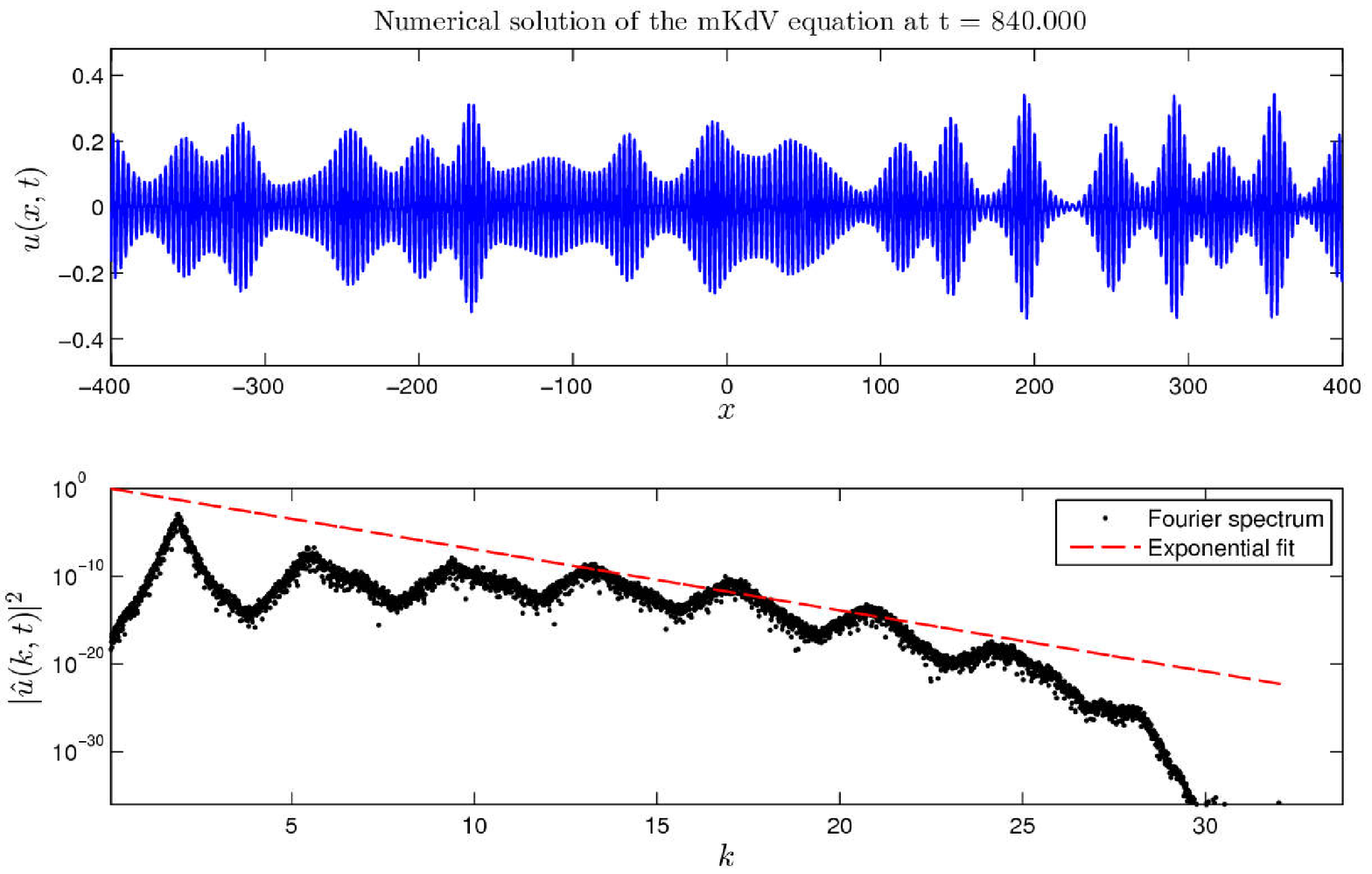}}
  \caption{\small\em Development of the \acs{mi} in the \acs{mkdv} equation for the initial wave amplitudes $a\ =\ 0.08$ (left panel) and $2a\ =\ 0.16$ (right panel) and several simulation times.}
  \label{fig:mi2}
\end{figure}

\begin{figure}
  \centering
  \includegraphics[width=0.99\textwidth]{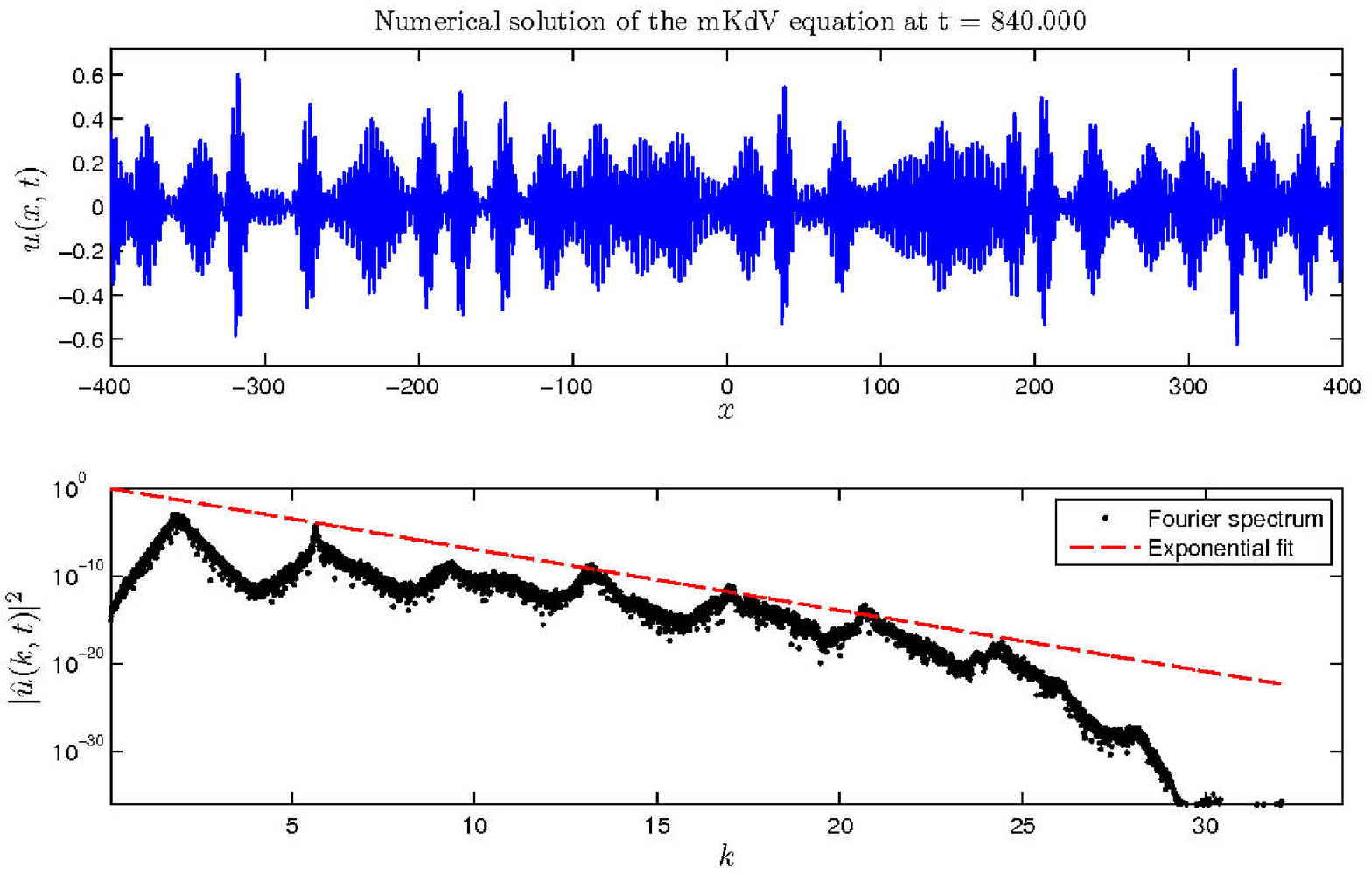}
  \caption{\small\em Development of the \acs{mi} in the \acs{mkdv} equation for the initial wave amplitude $3a\ =\ 0.24$ at the final simulation time $T\ =\ 840$.}
  \label{fig:mi3}
\end{figure}

In this Section we study the effect of the base wave amplitude on the development of the \acs{mi} in the \acs{mkdv} equation. Namely, we perform the same simulation described in the previous section with numerical parameters given in Table~\ref{tab:test1}, except for the amplitude $a$ which will be chosen as $2a\ =\ 0.16$ and $3a\ =\ 0.24$. Moreover, we will take a larger spectral domain $k\in[0, 32]$ in contrast to the previous case. A larger domain is precisely needed to observe several cascading modes and to make some conclusions on their distribution in the Fourier domain. The simulation results are shown on Figures~\ref{fig:mi2} and \ref{fig:mi3}.

In particular, one can see that the initial condition with the amplitude two times bigger ($a\ =\ 0.16$) develops the \acs{mi} much faster. According to the theoretical predictions \cite{Kartashova2013}, the time $T_{\mathrm{MI}}$ needed for the \acs{mi} to be fully developed in the physical space scales as $T_{\mathrm{MI}}\ \propto\ \O(\eps^{-2})$. So, the increase of the amplitude leads also to the increase in the nonlinearity $\eps$. Consequently, our numerical results corroborate this prediction of the theory. On the other hand, we stress that the position (and consequently the distance between) of the cascading modes remains unaffected by the change in the amplitude.

%%%%%%%%%%%%%%%%%%%%%%%%%%%%%%%%%%%%%%%%%%%%%%%%%%%%%%%%%%%%%%%%%%%%%%

\subsection{Effect of the excitation wavenumber}

\begin{figure}
  \centering
  \subfigure[$t\ =\ 48$]{
  \includegraphics[width=0.89\textwidth]{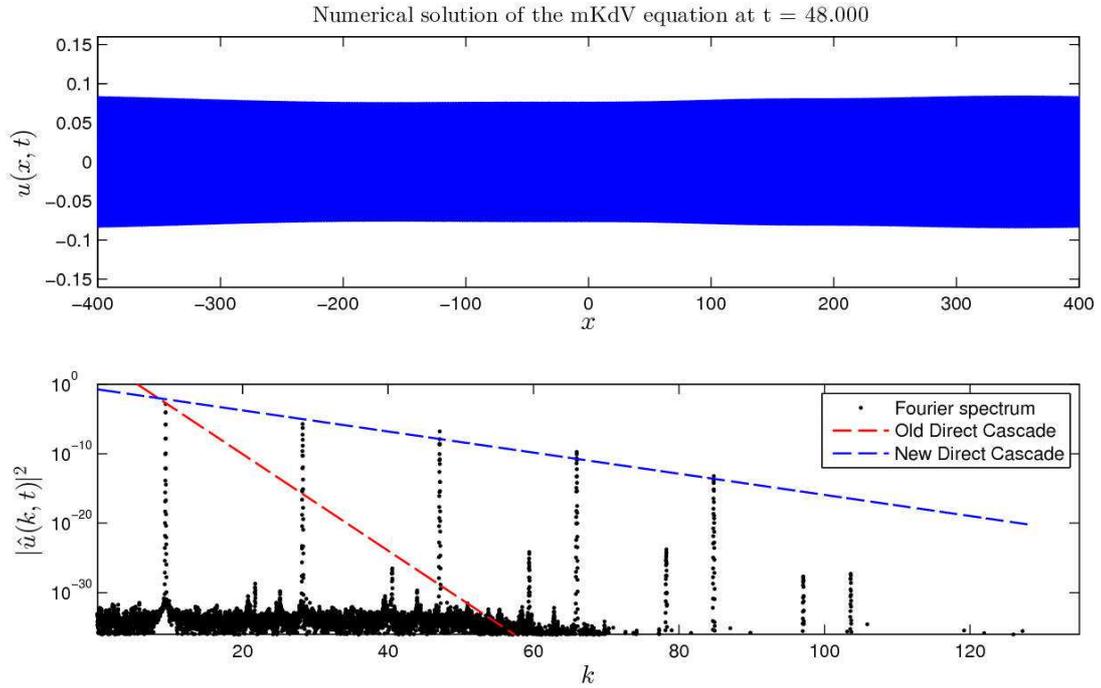}}
  \subfigure[$t\ =\ 150$]{
  \includegraphics[width=0.89\textwidth]{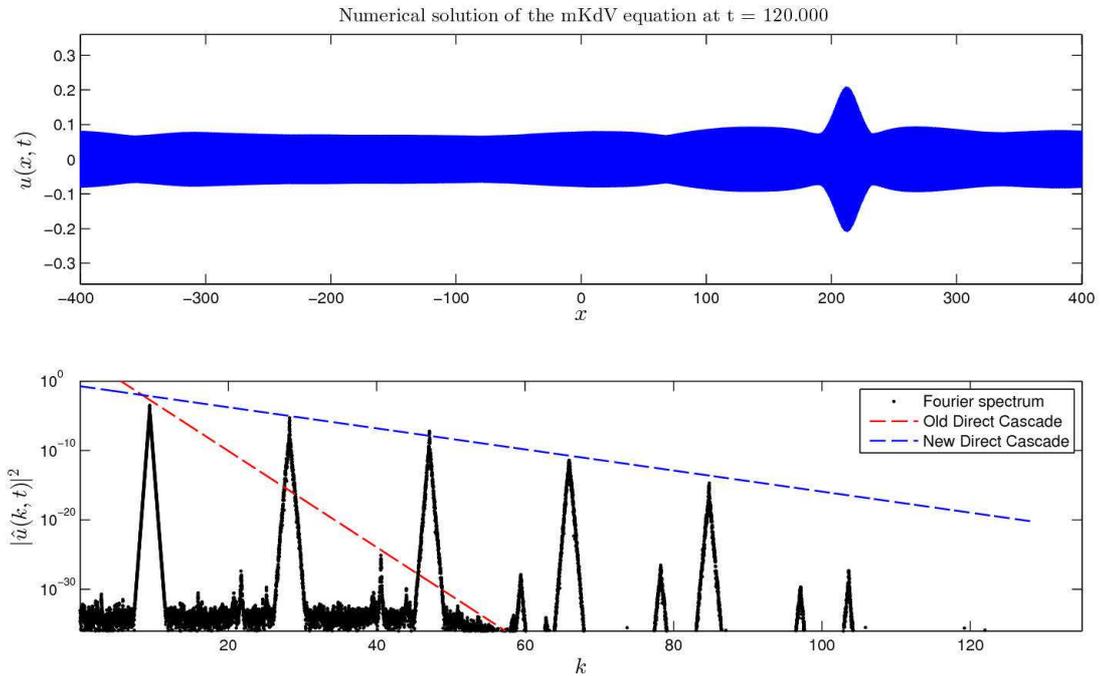}}
  \caption{\small\em \acs{mi} simulation for $k_0\ =\ 5\times 1.884$, all the other parameters are the same as given in Table~\ref{tab:test1}. The red dashed line on the bottom panels indicates the fit of the previous simulations for $k_0\ =\ 1.884$. The blue dashed line shows the new fit. The magenta dashed line shows the direction of the eventual inverse cascade.}
  \label{fig:k0dep}
\end{figure}

\begin{figure}
  \centering
  \subfigure[$a\ =\ 0.08$, $t\ =\ 115$]{
  \includegraphics[width=0.89\textwidth]{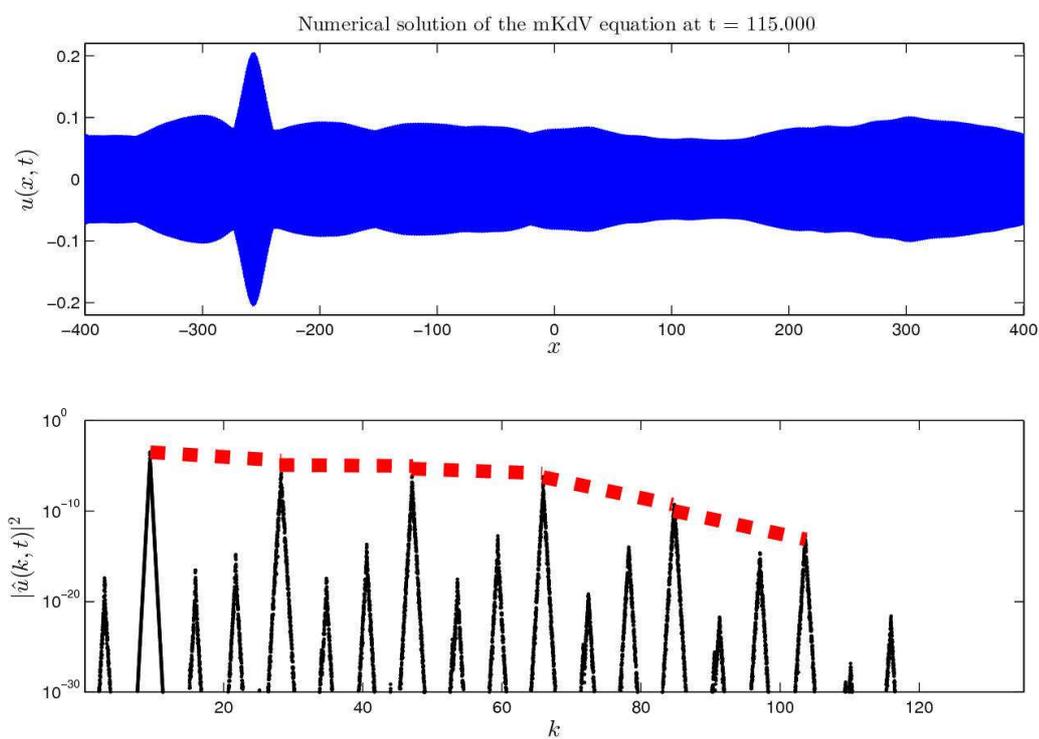}}
  \subfigure[$a\ =\ 0.16$, $t\ =\ 50$]{
  \includegraphics[width=0.89\textwidth]{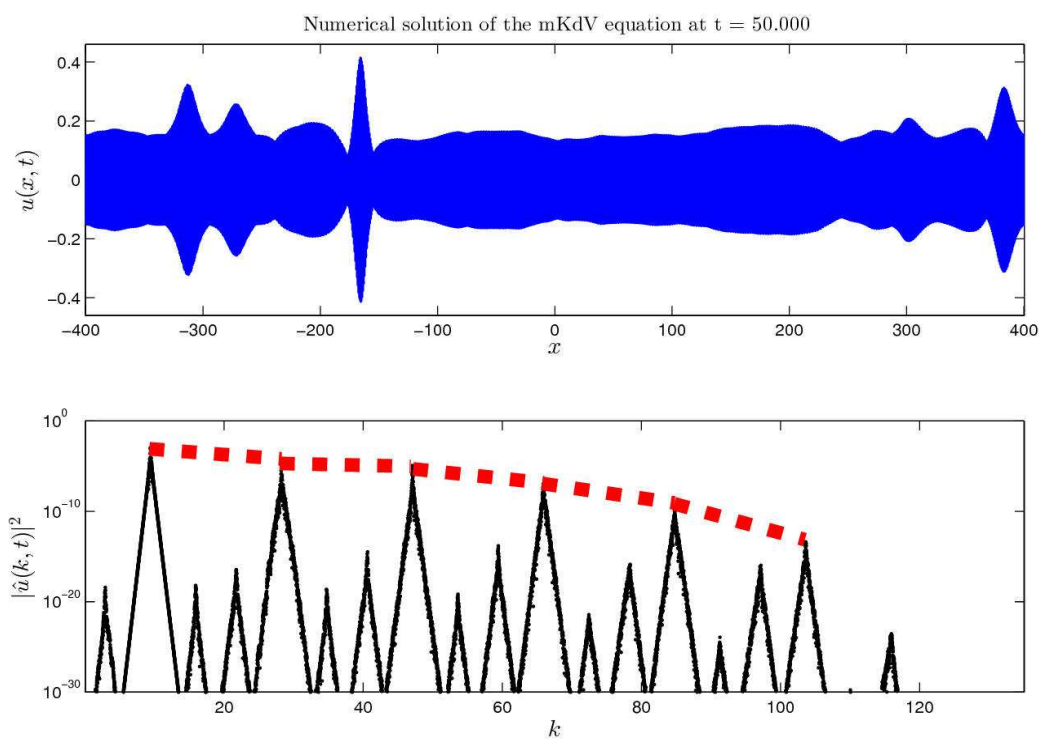}}
  \caption{\small\em \acs{mi} simulation for $k_0\ =\ 5\times 1.884$, and two different values of the base wave amplitude $a$. The red bold dotted line corresponds to the energy spectrum shape observed in the numerical simulation.}
  \label{fig:k0dep2}
\end{figure}

On Figures~\ref{fig:mi2} and \ref{fig:mi3} the base wavenumber $k_0$ was fixed and taken from Table~\ref{tab:test1}. In this Section we describe the numerical experiments for another value of the parameter $k_0\ =\ 5\times 1.884$. The results of numerical simulations are reported on Figures~\ref{fig:k0dep} and \ref{fig:k0dep2}. Namely, on Figure~\ref{fig:k0dep} it is shown that the new exponent of the spectrum is much lower than in the previous case. Moreover, if we increase the base wave amplitude $a$ for the new value of $k_0\ =\ 9.42$, the spectrum shape remains constant, however we cannot affirm anymore that it is still exponential as it is hinted on Figure~\ref{fig:k0dep2}. The positions and energies of cascading modes for these two simulations are reported in Table~\ref{tab:casc1}. From these observation one can clearly see that the initial wave amplitude does not affect the structure of the $D$-cascade.

Another important consequence of the variation of the parameter $k_0$ is the distance between cascading modes in the Fourier space. More precisely, the distance increases with the increase in $k_0$. For example, our simulations show that the distance increases from $\Delta k\ \approx\ 3.81$ to $18.42$ when $k_0$ goes from $1.884$ to $5\times 1.884\ =\ 9.42$.

\begin{table}
  \centering
  \begin{tabular}{|>{\columncolor[gray]{0.85}}l||>{\columncolor[gray]{0.85}}c||>{\columncolor[gray]{0.85}}c||>{\columncolor[gray]{0.85}}c||>{\columncolor[gray]{0.85}}c||>{\columncolor[gray]{0.85}}c||>{\columncolor[gray]{0.85}}l|}
  \hline\hline
    $a$ & $0.08$  &         &         &         &        &         \\
    \hline\hline
    $k$ & $9.428$ & $28.26$ & $47.11$ & $65.94$ & $84.8$ & $103.6$ \\
    \hline\hline
    $\E$ & $3.35\times 10^{-4}$ & $2.83\times 10^{-5}$ & $5.07\times 10^{-6}$ & $
    6.35\times 10^{-7}$ & $5.29\times 10^{-10}$ & $1.79\times 10^{-13}$ \\
    \hline\hline\hline
    $a$ & $0.16$  &         &         &         &        &         \\
    \hline\hline
    $k$ & $9.396$ & $28.27$ & $47.09$ & $65.92$ & $84.77$ & $103.6$ \\
    \hline\hline
    $\E$ & $1.04\times 10^{-3}$ & $4.30\times 10^{-5}$ & $1.18\times 10^{-5}$ & $8.15\times 10^{-8}$ & $3.52\times 10^{-10}$ & $3.2\times 10^{-14}$ \\
  \hline\hline
  \end{tabular}
  \bigskip
  \caption{\small\em Wavenumbers and energies of cascading modes for $a\ =\ 0.08$ and $a = 0.16$. The simulation snapshots are shown on Figure~\ref{fig:k0dep2}(\textit{a,b}) correspondingly.}
  \label{tab:casc1}
\end{table}

%%%%%%%%%%%%%%%%%%%%%%%%%%%%%%%%%%%%%%%%%%%%%%%%%%%%%%%%%%%%%%%%%%%%%%

\subsection{Effect of the perturbation magnitude}

\begin{figure}
  \centering
  \subfigure[$t\ =\ 5$]{
  \includegraphics[width=0.89\textwidth]{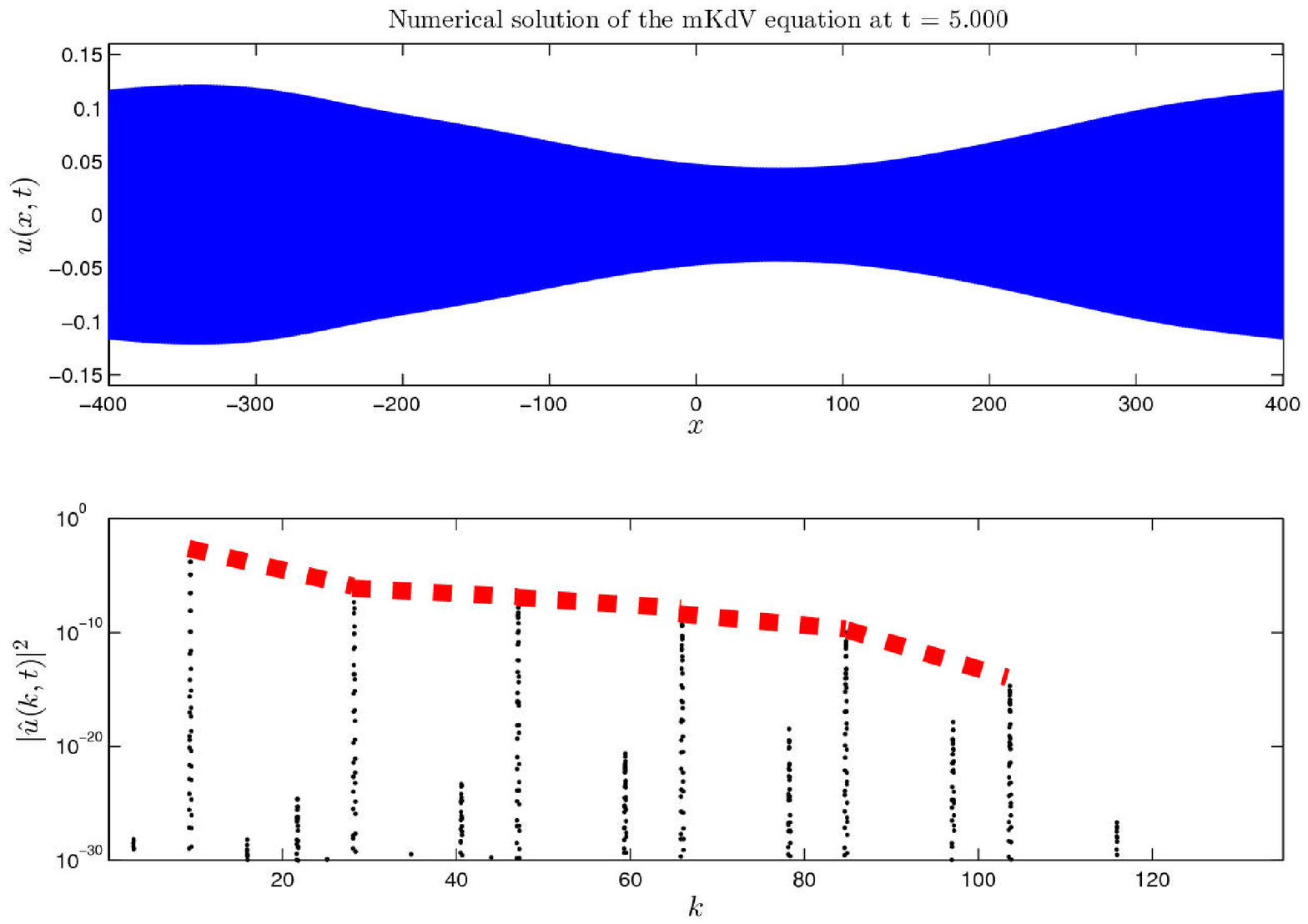}}
  \subfigure[$t\ =\ 40$]{
  \includegraphics[width=0.89\textwidth]{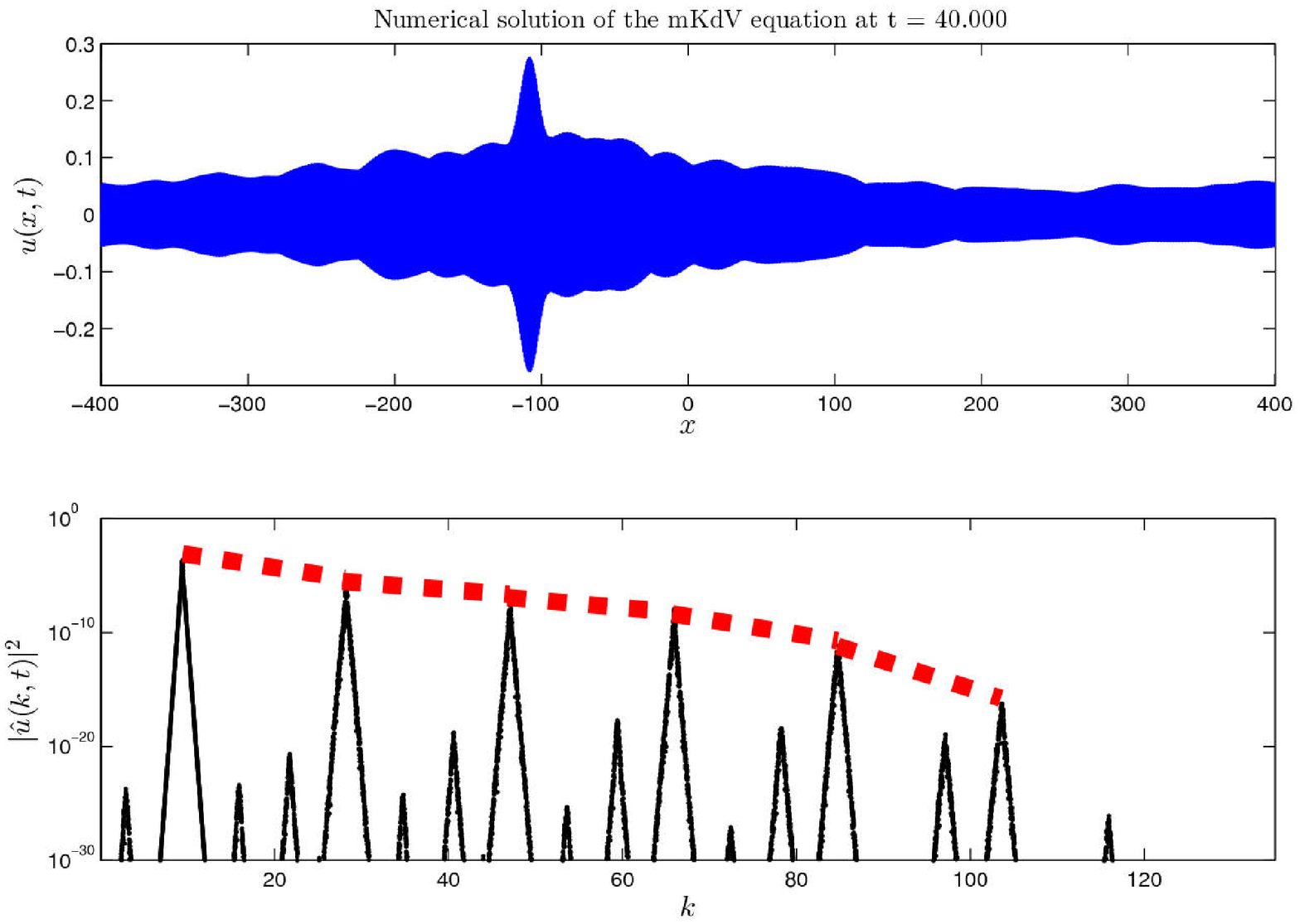}}
  \caption{\small\em \acs{mi} simulation for $k_0\ =\ 5\times 1.884$, and the perturbation magnitude $m\ =\ 10\times 0.05\ =\ 0.5$. The red bold dotted line corresponds to the energy spectrum shape observed in the numerical simulation.}
  \label{fig:m0dep}
\end{figure}

In the numerical simulations performed so far we have always taken the perturbation with magnitude $m\ =\ 0.05$, which results in 5\% amplitude modulation in terms of the base wave amplitude $a$. We tested various values of parameter $m$ in our numerical tests. Here we report on Figure~\ref{fig:m0dep} the most severe case of $m\ =\ 0.5$ (\ie 50\% of the perturbation). Even in such an extreme case, the $D$-cascade is still present. However, we cannot state anymore with certitude that its shape is exponential. This questions will require more detailed investigations with even more resolved numerical simulations.

%%%%%%%%%%%%%%%%%%%%%%%%%%%%%%%%%%%%%%%%%%%%%%%%%%%%%%%%%%%%%%%%%%%%%%

\section{Conclusions}\label{sec:disco}

In the present study we have investigated the $D$-cascade formation in the framework of the \acs{mkdv} equation. The main mechanism generating this cascade is the \acf{mi}. One of the first theoretical studies devoted to the \acs{mi} in the \acs{mkdv} equation is \cite{Grimshaw2001}, where the transformation from the \acs{mkdv} to the \acs{nls} equation was highlighted. This transformation allows to find the conditions when the \acs{mi} occurs in \acs{mkdv}. Their theoretical study was illustrated with some numerical simulations showing the development of the \acs{mi} in the physical space only. Since the $D$-model (or any other model) for energy transfer \emph{via} the \acs{mi} was not proposed yet at that time, the energy transport across the Fourier space was not in the focus. Consequently, the present study can be considered to be complementary since the main focus here is precisely on what is going on in the Fourier space.

In our numerical experiments we studied the nonlinear stage of the \acs{mi} evolution aiming to observe the evolution of Fourier spectra additionally to the wave observation in the physical space, already reported in previous studies \cite{Grimshaw2001, Ruderman2008}. The main findings can be briefly summarized in the following list:
\begin{itemize}
  \item For a wide class of the initial conditions leading to the \acs{mi} we clearly observe the formation of the direct (\ie in the direction of increasing wavenumbers $k$) $D$-cascade. The ranges of parameters considered in this study are given here:
  \begin{description}
    \item[Amplitude] $a\ =\ 0.01\ \div\ 0.24$
    \item[Perturbation magnitude] $m\ =\ 0.05\ \div\ 0.5$
    \item[Base wavenumber] $k_0\ =\ 1.8\ \div\ 60.0$
    \item[Simulation time horizon] $T\ =\ 40\ \div\ 2000\ \propto\ \O(\eps^{-2})$
    \item[Number of Fourier harmonics] $N\ =\ 1024\ \div\ 131\, 072$
  \end{description}

  \item In our numerical simulation the \acs{mi} develops on the dynamical time scales of the order of $\O(\eps^{-2})$ which is in agreement with theoretical predictions \cite{Kartashova2013}.

  \item It is interesting to note that this time scale refers to the complete development of the \acs{mi} in the physical space (see Figure~\ref{fig:mi2}(\textit{d, e})). On the other hand one can see that the main structure of the $D$-cascade is already observable in Fourier spectra from the first instances of the dynamical evolution. Consequently, the development of the \acs{mi} in the physical space corresponds to the spectral broadening of cascading modes. It is remarkable that their positions and energies are quasi-stationary.

  \item The observed $D$-cascade skeleton in the Fourier space is very robust and it has the exponential decay $\E_k\ \propto\ \exp(-\alpha \cdot k)$ (see Figures~\ref{fig:mi2} and \ref{fig:mi3}). The exponent $\alpha$ was found to be independent of the base wave amplitude $a$ for fixed values of other parameters.

  \item The increase of $k_0$ (the base wave wavenumber) has more drastic consequences on the spectrum shape. The agreement with the exponential spectrum $\E_k\ \propto\ \exp(-\alpha \cdot k)$ is not satisfactory anymore. This highly nonlinear regime requires a more detailed investigation with higher resolutions in the Fourier space in order to obtain a longer cascade before making some conclusions about its shape (most of simulations presented in this study were already performed with $N\ =\ 32\, 768$ Fourier harmonics).

  \item We performed a series of numerical experiments for different values of the perturbation magnitude $m = 0.05\ \div\ 0.5$. We conclude that the main features of the $D$-cascade described above are preserved (see Figure~\ref{fig:m0dep}). The choice of the perturbation influences the time scale on which the \acs{mi} will develop in the physical space. Generally, higher magnitudes of the perturbation tend to accelerate this process.
\end{itemize}

%%% ------------------------------------------------------------------------ %%%

\subsection*{Acknowledgments}
\addcontentsline{toc}{subsection}{Acknowledgments}

The authors would like to thank Professors Mat~\textsc{Johnson} (University of Kansas) and Fritz \textsc{Gesztesy} (University of Missouri) for interesting discussions and most useful comments. This research has been supported by the Austrian Science Foundation (FWF) under projects P22943-N18 and P24671.

%%% ------------------------------------------------------------------------ %%%

%%% Bibliography
\bigskip
\addcontentsline{toc}{section}{References}
\bibliographystyle{abbrv}
\bibliography{biblio}
\bigskip\bigskip

\end{document}